\documentclass[aps,prb,preprint,showpacs,superscriptaddress,longbibliography]{revtex4-1}
\usepackage{amsmath}
\usepackage{amsfonts}
\usepackage{amsbsy}
\usepackage{amssymb}
\usepackage{graphicx}
\usepackage{textcomp}
\usepackage[caption=false,singlelinecheck=false]{subfig}
\usepackage{xcolor}
\usepackage{mathrsfs}
\usepackage{mathtools}
\usepackage{bm}
\usepackage{gensymb}
\usepackage{braket,wasysym}
\usepackage[colorlinks,linkcolor=red,citecolor=red,filecolor=magenta,urlcolor=red,breaklinks]{hyperref}
\usepackage[bottom]{footmisc}
\usepackage{verbatim}

\hypersetup{colorlinks=true, urlcolor=blue, citecolor=magenta, pdfborder={0 0 0}}
\usepackage{breakurl}
\usepackage{natbib}

\allowdisplaybreaks

\begin{document}
\title{Discovery of new quasicrystals from translation of hypercubic lattice
}

\author{Junmo Jeon}
\email{junmo1996@kaist.ac.kr}
\affiliation{Korea Advanced Institute of Science and  Technology, Daejeon 34141, South Korea}
\author{SungBin Lee}
\email{sungbin@kaist.ac.kr}
\affiliation{Korea Advanced Institute of Science and  Technology, Daejeon 34141, South Korea}

\date{\today}
\begin{abstract}
How are the symmetries encoded in quasicrystals? As a compensation for the lack of translational symmetry, quasicrystals admit non-crystallographic symmetries such as 5- and 8-fold rotations in two-dimensional space. It is originated from the extended crystallography group of high-dimensional lattices and projection onto the physical space having a finite perpendicular window. One intriguing question is : How does the translation operation in high-dimension affect to the quasicrystals as a consequence of the projection? Here, we answer to this question and prove that a simple translation in four-dimensional hypercubic lattice completely determines a distinct rotational symmetry of two-dimensional quasicrystals. In details, by classifying translations in four-dimensional hypercubic lattice, new types of quasicrystals with 2,4 and 8-fold rotational symmetries are discussed. It gives us an important insight that a translation in high-dimensional lattice is intertwined with a rotational symmetry of the  quasicrystals. 
{In addition, regarding to the multi-frequency driven systems, it provides a unique way to discover new quasicrystals projected from the high-dimensional Floquet lattice.}
\end{abstract}
\maketitle

\section{Introduction}
\label{sec: intro}

The symmetries of the lattice are important along with a broad impact on the physical phenomena\cite{chaikin1995principles,tsvelik2007quantum,marder2010condensed,simon2013oxford,jones1985theoretical,cohen2016fundamentals,de2012structure,haldane1985many,bernard1991quantum,kittel1976introduction,estrada2009influence,petsas1994crystallography,macchesney1965electric,dmitriev2005symmetry,khan2019cubic}. Fortunately, the crystallography group, the set of symmetries compatible with the lattice translational symmetry, has been clarified for any dimensions\cite{de2012structure,petsas1994crystallography,yamamoto1996crystallography,baake2017aperiodic,walter2009crystallography,kittel1976introduction,cohen2016fundamentals}. In particular, the two-dimensional lattices allow 2,3,4 and 6-fold rotations but forbid 5, 7 and 8-fold rotations for example\cite{cohen2016fundamentals,de2012structure}. Thus, the crystallography group plays a crucial role in understanding crystal structures and their physical properties\cite{petsas1994crystallography,yamamoto1996crystallography,baake2017aperiodic,walter2009crystallography,kittel1976introduction,cohen2016fundamentals,fuentes2004magnetoelectricity,das2020crystallographic,de2012structure,routledge2021mode,perez2015symmetry}. Meanwhile, the quasicrystalline lattices, ordered but no translational symmetry, do not belong to these crystallography group\cite{cohen2016fundamentals,walter2009crystallography}.  For instance, there are 5- and 8-fold rotationally symmetric quasicrystals in two dimension, known as Penrose tiling and Ammann-Beenker tiling respectively\cite{PhysRevLett.59.1010,bursill1985penrose}. 
Despite the lack of translational symmetry, it compensates the rich physical phenomenon such as critical states and phason modes, which are uniquely present in quasicrystals\cite{PhysRevResearch.3.013168,jeon2020phonon,jeon2021length,jeon2021pattern,stadnik2012physical,schwabe1999influence,kohmoto1987critical,mace2017critical,vekilov2000influence,bandres2016topological,freedman2006wave,yu2019dodecagonal,cryst8110416,arai1988strictly,kim2007novel,oktel2021strictly,cryst8100370,cryst7100304,de2012phonons,homes1991optical,de2011phonons,bistritzer2011moire,nemec2007hofstadter}. 
Thus, many researchers have searched for the extended rule to control their symmetries in terms of the symmetry classification in quasicrystalline lattices\cite{pezzini202030,reinhardt2013computing,noya2021design,reinhardt2016self,tracey2021programming,yu2019dodecagonal,oktel2021strictly,cryst8100370,cryst7100304,de2012phonons,homes1991optical,de2011phonons}. It gives us not only the insight of the currently existing quasicrystal structures but also the opportunity to find new quasi-periodic tiling patterns. 
%
%

To understand non-crystallographic symmetries of quasicrystals, one way is to look for the crystallography group of higher-dimension and their projection onto the lower-dimension\cite{senechal1996quasicrystals,petsas1994crystallography,yamamoto1996crystallography,baake2017aperiodic,walter2009crystallography}. In general, higher dimension allows extended crystallography group\cite{cohen2016fundamentals}. For instance, 5-fold and 8-fold rotational symmetries are present in the five-dimensional and four dimensional crystallography groups, respectively. Thus, octagonal quasicrystal such as an Ammann-Beenker tiling, can be built by projecting 8-fold rotational symmetric lattice points in four dimension into the two-dimensional subspace. The process explained above is often called as the cut-and-project scheme (CPS), the most well known method\cite{aragon2019twisted,petsas1994crystallography,yamamoto1996crystallography,baake2017aperiodic,walter2009crystallography}. 
Based on the CPS, it is already known that distinct high-dimensional lattice structures such as face-centered-cubic and body-centered-cubic result in different quasicrystals\cite{indelicato2012structural}. However, the fundamental question -- what kinds of operation in high-dimensional lattice are intertwined with the one in low-dimensional quasicrystals, remains yet to be investigated. In detail, beyond the phason dynamics, it is required to deepen our understanding of quasicrystals and their symmetries from the particular operations in high-dimensional lattice, such as a simple translation. 


Nowadays, many quasicrystalline lattices are experimentally accessible in the optical lattice, using multiple lasers that play a role of high-dimensional degree of freedoms\cite{jagannathan2014eightfold,corcovilos2019two,sanchez2005bose,viebahn2019matter,sbroscia2020observing}. In the optical lattice, the quasicrystalline potential gradient pattern could be built by the interference of the multiple lasers\cite{viebahn2019matter,sbroscia2020observing}. Particularly, the 8-fold rotational symmetric quasicrystalline potential for ultracold atoms has been experimentally demonstrated\cite{viebahn2019matter,sbroscia2020observing}.
Even in the biological systems, engineering icosahedral symmetric quasicrystalline polymers such as DNA and viral capside have been discussed\cite{reinhardt2016self,GRIMM2003731,noya2021design}. 
Moreover, multi-frequency driven systems are in the spotlight as the candidate of the higher-dimensional lattices\cite{zhao2013hamiltonian,russomanno2017floquet,martin2017topological,rodriguez2021real,malz2021topological,poertner2020validity,thomson2021flow}. It has been realized making time translation symmetry being discretized by  multiple periodic drivings and resulting in the Floquet lattice systems\cite{martin2017topological,rodriguez2021real,malz2021topological}. Thus, one could also expect to  manipulate a translation operation in high-dimensional systems and find new types of quasicrystals.


In this paper, we show how the translation operation in high dimensional lattice 
is intertwined with the symmetry of quasicrystals, when they are projected onto the physical space. In particular, we focus on quasicrystals projected from four dimensional hypercubic lattice. 
Surprisingly, despite the lost of translation symmetry from projection, the translation of the high dimensional lattice is embedded in the rotation symmetry of quasicrystals. It turns out that a translation of four dimensional hypercubic lattice completely changes the rotation symmetry of two dimensional quasicrystals. 
In detail, the special set of translations in high-dimensional lattice enables modification of the quasi-periodic tiling pattern and its symmetry. Exemplifying the translation of four-dimensional hypercubic lattice, we illustrate the formation of new quasicrystals beyond the most well known Ammann-Beenker tiling. 
Moreover, such new quasicrystalline structures have different orders of the rotational symmetries, depending on the translation vector.
Our works offers a new insight of quasicrystalline symmetries that are intertwined with the translation of the high-dimensional lattice. Furthermore, it suggests the novel way to generate new types of quasicrystals in both tiling pattern and symmetry aspects. 
\begin{figure}[]
\centering
\includegraphics[width=0.9\textwidth]{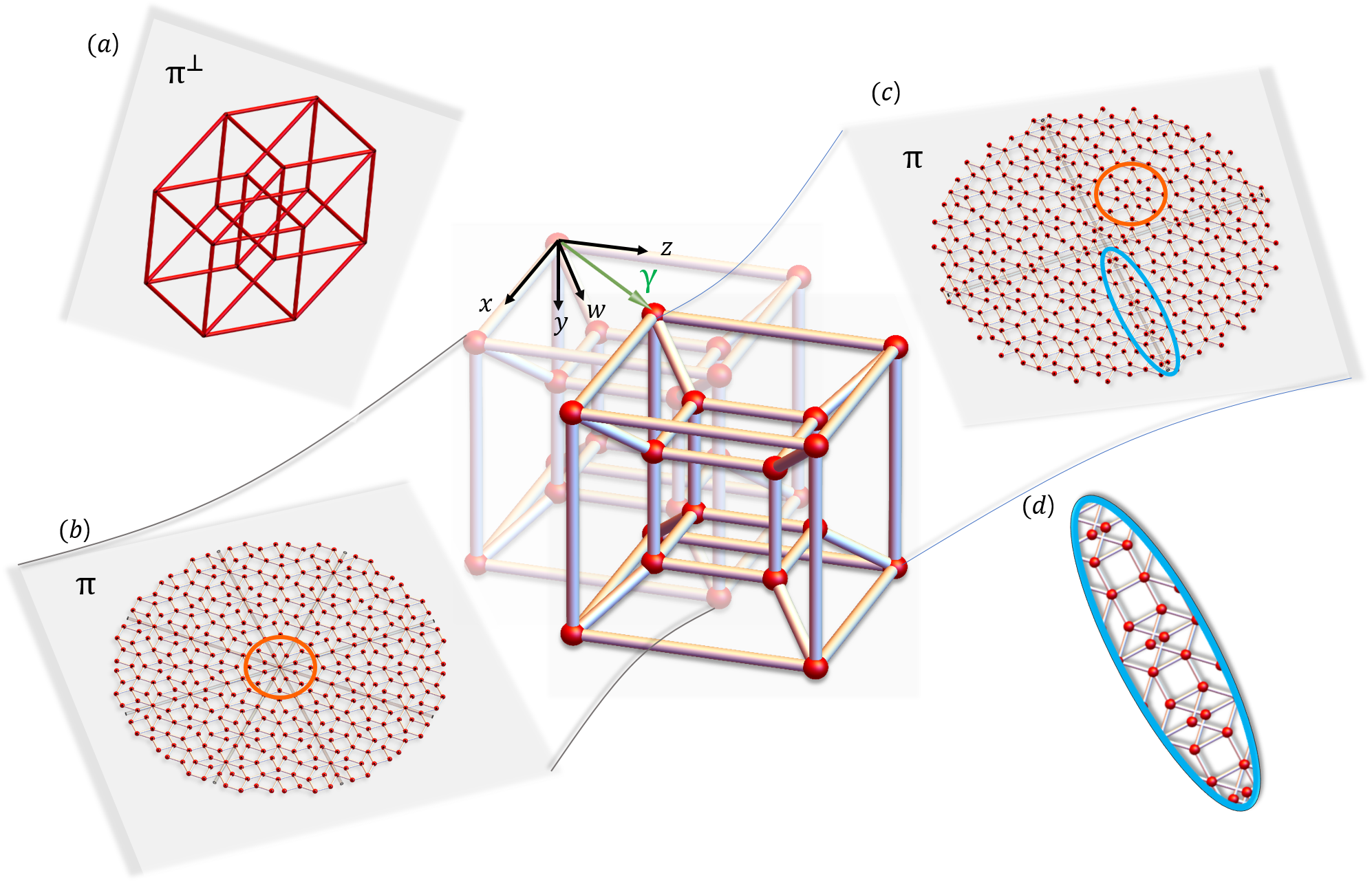}
\label{fig1}
\caption{\label{fig: window} Schematic illustration of finding new quasicrystals from a translation operation in the high-dimensional lattice. The unit cell of four-dimensional hypercubic lattice -- a tesseract, is drawn at the center. 
(The faint image in the back is a tesseract before translation with a shifting vector $\gamma$.) (a) The window formed by the $\boldsymbol{\pi}^\perp$-projection of the Wigner-Seitz cell.
The special set of translations given by the shift vector $\boldsymbol{\gamma}$ (green arrow) of the high-dimensional lattice generates new types of quasicrystals. Before translation, the $\boldsymbol{\pi}$-projected quasicrystal is the 8-fold symmetric Ammann-Beenker tiling. (c) With translation by  $\boldsymbol{\gamma}=(x,y,z,w)=\frac{1}{2}(1,0,1,0)$, the $\boldsymbol{\pi}$-projection results in a new type of quasicrystal with the 4-fold symmetry. 
In both (b) and (c), the local 8-fold symmetry is shown as marked in orange circle. The gray lines are drawn for guide to the eyes to emphasize 8-, 4-fold symmetries, respectively. 
(d) Enlarged image of the new local tiling pattern shown as a blue circle in (c), which is absent in the Ammann-Beenker tiling (b).
}
\end{figure}


\section{Cut-and-project scheme (CPS)}
\label{sec: cps}
Before discussing the intertwined quasicrystal symmetries and translation of high-dimensional lattices, let us briefly review the cut-and-project scheme (CPS).
 As introduced above, the CPS is a way to understand non-crystallographic symmetry of the low-dimensional quasicrystals in terms of the  high-dimensional lattices\cite{senechal1996quasicrystals}.
We first consider high-dimensional cubic lattice $\mathcal{L}=\{ \boldsymbol{x} | \boldsymbol{x}=m_i \boldsymbol{e}_i, m_i\in\mathbb{Z}, 1\le i\le D\}$, where $D$ is the dimension of the lattice and $\boldsymbol{e}_i$ is a standard unit vector. The $D$-dimensional space is decomposed into two projection maps $\boldsymbol{\pi}$ and $\boldsymbol{\pi}^{\perp}$. Each of them projects the lattice points in $\mathcal{L}$ onto the subspace of the quasicrystal (physical space) and its orthogonal complement subspace (perpendicular space), respectively. 
 To produce the nontrivial quasicrystalline pattern, the physical space should have an irrational angle to the lattice surface. 
However, for such an irrational angle, the images of the projection of entire lattice points, $\boldsymbol{\pi}(\mathcal{L})$, densely cover the physical space. 
Thus, one should select a subset of $\mathcal{L}$, which is the compact subset of the perpendicular space $\boldsymbol{\pi}^{\perp}$ and is often termed as the window, say $K$.
 Now, we project the lattice points only when those image of $\boldsymbol{\pi}^{\perp}$ belongs to the window $K$. Then, the resultant projection image onto the physical space emerges the discrete quasicrystalline lattice structure. As a standard choice of the window, $K=\boldsymbol{\pi}^{\perp}(\mathcal{W}(0))$, where, $\mathcal{W}(0)$ is the Wigner-Seitz cell of the origin\cite{senechal1996quasicrystals,cryst8100370,cryst7100304}.

To illustrate the details, let us consider the example of the CPS, the Ammann-Beenker tiling. The Ammann-Beenker tiling is generated by the CPS with four-dimensional hyper-cubic lattice with $D=4$. The Wigner-Seitz cell is given by the polytope whose 16 verticies are $(\pm\frac{1}{2},\pm\frac{1}{2},\pm\frac{1}{2},\pm\frac{1}{2})$ where the signs of each component are independent. 
The $\boldsymbol{\pi}$ and $\boldsymbol{\pi}^{\perp}$ projections are given by,
\begin{align}
\label{projection}
\boldsymbol{\pi} &=\begin{pmatrix} 1 & \frac{1}{\sqrt{2}} & 0 & -\frac{1}{\sqrt{2}} \\ 0 & \frac{1}{\sqrt{2}} & 1 & \frac{1}{\sqrt{2}} \end{pmatrix} \\
\label{projectionperp}
\boldsymbol{\pi}^{\perp}&=\begin{pmatrix} 1 & -\frac{1}{\sqrt{2}} & 0 & \frac{1}{\sqrt{2}} \\ 0 & \frac{1}{\sqrt{2}} & -1 & \frac{1}{\sqrt{2}} \end{pmatrix}.
\end{align}
Then, the window $K$ becomes a tesseract graph whose vertices are $\boldsymbol{\pi}^{\perp}$-projection images of the vertices of the Wigner-Seitz cell (See Fig.\ref{fig: window} (a).)\cite{baake2017aperiodic}. The Ammann-Beenker tiling (See Fig.\ref{fig: window} (c)) is given by the $\boldsymbol{\pi}$-projection of four-dimensional hyper-cubic lattice points, whose image of $\boldsymbol{\pi}^{\perp}$ belongs to the window $K$ in Fig.\ref{fig: window} (a).

\section{New quasicrystals from high-dimensional translation}
\label{sec: new}

Now, we discuss the quasicrystalline symmetry and tiling pattern originated from the translation operation of the high-dimensional lattice.  
Since the quasicrystal is formed via the projection, one would immediately suspect that the change in the projection modifies the quasicrystalline structure. Similarly, if we enlarge or shear the window, the resultant quasicrystal could be distinct as well\cite{indelicato2012structural}. Nevertheless, the explicit relationship between high-dimensional lattice operation and formation of the quasicrystal remains barely understood, which is a prominent step to study non-crystallographic quasicrystalline tiling patterns\cite{cryst8100370,cryst7100304}. Specifically, it clarifies how the tiling patterns and the symmetries of the quasicrystals are descended from high-dimensional lattice  operation such as a simple translation. Hence, the construction of the complex quasicrystals can become controllable in terms of the high-dimensional lattice operation.
In such a vein, it has been required to find a systematic way of using high-dimensional lattice operation to generate distinct quasicrystals beyond the existing methods.

We consider the high-dimensional simple cubic lattice whose lattice points are given by $\mathbb{Z}^D$. Now, let's make a translation of the high-dimensional cubic lattice by the shift vector, $\boldsymbol{\gamma}$. Then, the high-dimensional lattice points are shifted as $\mathbb{Z}^D+\boldsymbol{\gamma}$. 
Although there exists an arbitrary choice of the shift vector, $\boldsymbol{\gamma}$, it turns out that the resultant quasicrystal pattern and its symmetry do not drastically change for all but for some special sets of $\boldsymbol{\gamma}$. For random $\boldsymbol{\gamma}$, in general, the local tiling patterns of the quasicrystals remain unchanged, but the global symmetry of the quasicrystal is being eliminated. On the other hand, for the special sets of $\boldsymbol{\gamma}$, a new local tiling pattern of the quasicrystal emerges along with the rotational symmetry. Thus, we mainly focus on these special sets of $\boldsymbol{\gamma}$ and study how the rotational symmetry of quasicrystals are changed, resulting in distinct and new quasi-periodic tiling patterns. 

To explore it, we exemplify a four dimensional hypercubic lattice described by the lattice points $\boldsymbol{r}_{4D} =  (m,n,l,k)^T$, where $m,n,l,k$ are integers. Here, $T$ stands for transposition.
By applying Eq.\eqref{projection} and Eq.\eqref{projectionperp}, we get the projection images of $(m,n,l,k)$ given by $\boldsymbol{\pi} \circ \boldsymbol{r}_{4D}=\left(m+\frac{1}{\sqrt{2}}(n-k),l+\frac{1}{\sqrt{2}}(n+k)\right)$ and $\boldsymbol{\pi}^{\perp} \circ \boldsymbol{r}_{4D}=\left(m-\frac{1}{\sqrt{2}}(n-k),-l+\frac{1}{\sqrt{2}}(n+k)\right)$, respectively. 
The quasicrystalline lattice points are comprised of $\boldsymbol{\pi} \circ \boldsymbol{r}_{4D}$ only if $\boldsymbol{\pi}^\perp  \circ \boldsymbol{r}_{4D}$ belongs to the window shown in Fig.\ref{fig: window} (a).
In this case, as already known, the resultant quasicrystal is the Ammann-Beenker tiling, which has an $8$-fold rotational symmetry. Specifically, the tiling has both local and global $8$-fold rotational symmetry (See Fig.\ref{fig: window} (b)). 

Now, we discuss the case that a four dimensional hypercubic lattice is translated by a shift vector, $\boldsymbol{\gamma}$.
Although the global rotational symmetry is generally absent with respect to the random translations, surprisingly, the special sets of $\boldsymbol{\gamma}$ allow the global rotational symmetries such as 2,4 and 8-fold.
 Moreover, it results in new quasi-periodic tiling patterns.
To illustrate the idea,  
let us consider $\boldsymbol{\gamma}=\frac{1}{2}(0,1,0,1)^T$, as an example. This shift vector translates $n$ and $k$. Then, the projection images are given by,
$\boldsymbol{\pi} \circ \big( \boldsymbol{r}_{4D} + \boldsymbol{\gamma} \big)=\left(m+\frac{1}{\sqrt{2}}(n-k),l+\frac{1}{\sqrt{2}}(n+k+1)\right)$ and 
$\boldsymbol{\pi}^{\perp} \circ \big( \boldsymbol{r}_{4D} + \boldsymbol{\gamma} \big)=\left(m-\frac{1}{\sqrt{2}}(n-k),-l+\frac{1}{\sqrt{2}}(n+k+1)\right)$.
We claim that these new quasicrystalline lattice points have the $4$-fold global rotational symmetry rather than $8$-fold in addition to the distinct local pattern. 
Note that the $8$-fold and $4$-fold rotation in two-dimensional space is given by the rotation matrices, 
$ R_8=\frac{1}{\sqrt{2}}\begin{pmatrix} 1 & -1 \\ 1 & 1 \end{pmatrix} $ and $R_4=(R_8)^2$, respectively.  
By applying $R_8$ to $\boldsymbol{\pi} \circ \big(\boldsymbol{r}_{4D} + \boldsymbol{\gamma} \big) $, one can easily find it does not belong to any lattice points $\boldsymbol{\pi} \circ \big(\boldsymbol{r}_{4D} + \boldsymbol{\gamma} \big) $
, indicating that new quasicrystalline lattice do not contain the $8$-fold rotational symmetry.
Instead, the new quasicrystalline lattice has the $4$-fold rotation symmetry, where one can find $\boldsymbol{r'}_{4D}$ that satisfies 
$\boldsymbol{\pi} \circ \big( \boldsymbol{r'}_{4D} + \boldsymbol{\gamma} \big) = R_4 \circ \boldsymbol{\pi} \circ \big( \boldsymbol{r}_{4D} + \boldsymbol{\gamma} \big)$.
One can also show that the points projected to the perpendicular space also satisfy the $4$-fold symmetry.
Thus, the new quasicrystal generated by the shift vector $\boldsymbol{\gamma}=\frac{1}{2}(0,1,0,1)^T$ has a global $4$-fold rotational symmetry rather than an $8$-fold one.

Fig.\ref{fig: window} illustrates a schematic diagram of high dimensional lattice translation and their resultant quasicrystals. The four-dimensional hypercubic lattice is projected onto $\boldsymbol{\pi}$-space with a finite window $\boldsymbol{\pi}^\perp$ shown in Fig.\ref{fig: window} (a). Before the translation, the $\boldsymbol{\pi}$-projection gives rise to the Ammann-Beenker tiling. A new quasiperiodic pattern in Fig.\ref{fig: window} (c)  emerges due to the translation of the hypercubic lattice $\boldsymbol{\gamma}=\frac{1}{2}(0,1,0,1)^T$.
As marked in orange circles in Figs.\ref{fig: window} (b) and (c), the local $8$-fold pattern is present in both the Ammann-Beenker tiling and the new quasicrystal pattern.
 However, unlike the Ammann-Beenker case, the global rotational symmetry is now a 4-fold not an 8-fold. (The gray lines in Fig.\ref{fig: window} (b) and (c) are drawn for guide to the eyes, to emphasize the 4-fold and 8-fold symmetries, respectively.)
 In addition, a new quasi-periodic pattern particularly along the diagonal lines emerges (See Fig.\ref{fig: window} (d)).
  It extends along the diagonal lines in the thermodynamic limit, being comprised of local $4$- and $2$-fold symmetric patterns but no $8$-fold symmetric pattern.

\section{Commensurate translation and symmetry of quasicrystalline lattice}
\label{sec: sym}
The specific example in Sec.\ref{sec: new} implies that a simple translation operation in the hypercubic lattice gives rise to the new quasicrystal with different rotational symmetry. 
 Now, one could ask what generic conditions of $\boldsymbol{\gamma}$ are required for distinct quasicrystalline symmetries. To explore it, we take into account the general case with the commensurate translations of the hypercubic lattice, where all components of $\boldsymbol{\gamma}$ are the rational numbers.

\begin{figure*}[]
	\centering
	\includegraphics[width=0.8\textwidth]{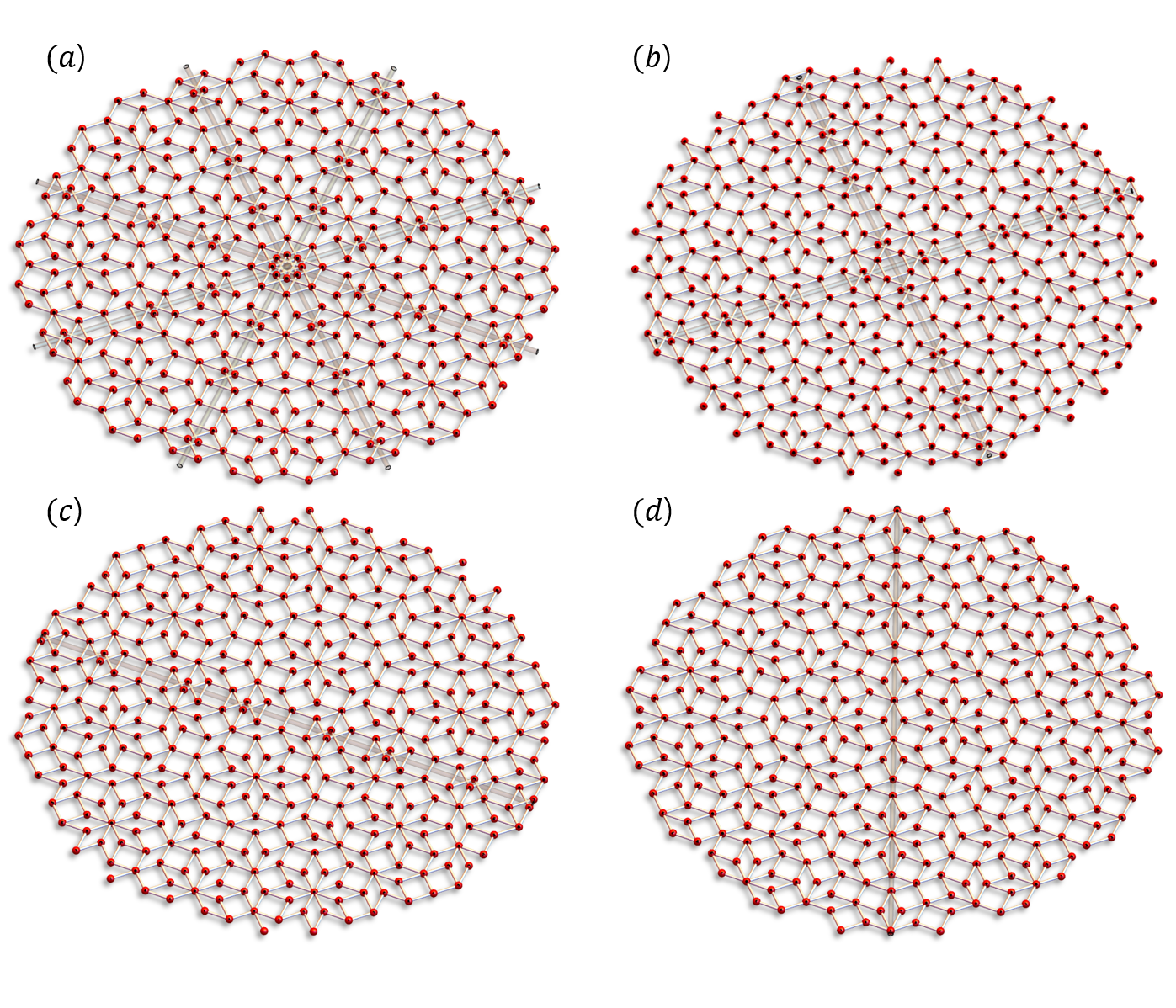}
	\label{fig2}
	\caption{\label{fig: summary} New quasicrystalline patterns with distinct rotational symmetries originated from the translation of hypercubic lattice. 
	(a) 8-fold rotational symmetric pattern characterized by the translation $\boldsymbol{\gamma}=\frac{1}{2}(1,1,1,1)^T$,  (b) 4-fold pattern with $\boldsymbol{\gamma}=\frac{1}{2}(0,1,0,1)^T$, (c) 2-fold pattern with $\boldsymbol{\gamma}=\frac{1}{2}(1,0,0,0)^T$ (d) 2-fold pattern with $\boldsymbol{\gamma}=\frac{1}{2}(1,1,0,0)^T$.
	 (a)-(c) Along the diagonal lines, a new quasi-periodic pattern shown in Fig.\ref{fig: window} (d) emerges, similar to the case of Fig.\ref{fig: window} (c). The gray lines in (a)-(d) are drawn for guide to the eyes, to emphasize 8, 4 and 2-fold symmetries, respectively.
	}
\end{figure*}

Let the shift vector, $\boldsymbol{\gamma}=(\gamma_m,\gamma_n,\gamma_l,\gamma_k)^T$, where $\gamma_m,\gamma_n,\gamma_l$ and $\gamma_k$ are all rational numbers. 
Note that the four-dimensional lattice points are given by $\mathbb{Z}^4$, thus it is enough to consider a modulo 1 of the components of $\boldsymbol{\gamma}$. The four-dimensional lattice points shifted by $\gamma$, are projected onto the physical space, $\pi$-projection, 
\begin{widetext}
\begin{align}
\label{physicalgen}
&\boldsymbol{\pi} \circ \big( \boldsymbol{r}_{4D} + \boldsymbol{\gamma} \big)=\left(m+\gamma_m+\frac{1}{\sqrt{2}}(n-k+\gamma_n-\gamma_k),l+\gamma_l+\frac{1}{\sqrt{2}}(n+k+\gamma_n+\gamma_k)\right).
\end{align}
\end{widetext}
In order to have either 4-fold or 8-fold symmetries, the rotated points of Eq.\eqref{physicalgen} by applying $R_4$ and $R_8$ should belong to the lattice points, and this leads to the following conditions,
\begin{align}
\label{qudra}
&(\gamma_m\!+\!\gamma_l, \gamma_m\!-\!\gamma_l, \gamma_k\!+\!\gamma_n, \gamma_k\!-\!\gamma_n) \in\mathbb{Z}^4 \mbox{ for 4-fold}\\
\label{octa}
&(\gamma_k\!+\!\gamma_m, \gamma_n\!-\!\gamma_l, \gamma_m\!-\!\gamma_n, \gamma_l\!-\!\gamma_k) \in\mathbb{Z}^4 \mbox{ for 8-fold.}
\end{align}
Based on Eq.\eqref{qudra}, the $4$-fold symmetry is present if 
$(\gamma_m,\gamma_l)$ and $(\gamma_n,\gamma_k)$ are either $(0,0)$ or $\frac{1}{2}(1,1)$. Particularly, the specified example in Sec.\ref{sec: new} corresponds to $(\gamma_m, \gamma_l)=(0,0)$ and $(\gamma_n, \gamma_k)=\frac{1}{2}(1,1)$
i.e., $\boldsymbol{\gamma} = (0,\frac{1}{2},0, \frac{1}{2})^T$. On the other hand, for the $8$-fold symmetry, Eq.\eqref{octa} indicates the presence of quasicrystals if and only if either $\boldsymbol{\gamma}=(0,0,0,0)^T$ or $\boldsymbol{\gamma}=\frac{1}{2}(1,1,1,1)^T$. Thus, there are only two distinct 8-fold quasicrystals for $\boldsymbol{\gamma}=(0,0,0,0)^T$ (Ammann-Beenker tiling) and $\boldsymbol{\gamma}=\frac{1}{2}(1,1,1,1)^T$ (See Fig.\ref{fig: summary} (a).), respectively.

Fig.\ref{fig: summary} illustrates the symmetric quasicrystalline lattices originated from various high-dimensional commensurate translations. Fig.\ref{fig: summary} (a) and (b) represent 8-, and 4-fold rotational symmetric cases, respectively, while Figs.\ref{fig: summary} (c) and (d) represent 2-fold symmetric cases. It demonstrates that the rotational symmetry of the quasicrystalline lattice is controllable in terms of the shift vector, $\boldsymbol{\gamma}$. In addition, Figs.\ref{fig: summary} (a)-(c) also show anomalous new local pattern along the diagonals as introduced in Fig.\ref{fig: window} (d), similar to the case of Fig.\ref{fig: window} (c).
Note that such new local pattern is absent in the case of Fig.\ref{fig: summary} (d). 
(See Supplementary Material for details of Fig.\ref{fig: summary} with their $\boldsymbol{\pi}^\perp$-projection images.) 
Table.\ref{table: summary} shows the classification of distinct rotational symmetries and their quasi-periodic patterns for general commensurate translations.
In details, there are six different classes of the shift vectors for the four dimensional hypercubic lattice. (See Supplementary Material for the asymmetric tiling from general commensurate translation.) \begin{table*}[]
\begin{tabular}{c|ccccc}
 \hline
                                                 & \multicolumn{5}{c}{$n$-fold Symmetric}                                                                                                                                                                                                                                                                                                                                                                                                                                                                                                                                                                     \\ \hline \hline
$n$                                               & \multicolumn{2}{c|}{8}                                                                                                                                                                        & \multicolumn{1}{c|}{4}                                                                                                                                  & \multicolumn{2}{c}{2}                                                                                                                                                                                                                            \\ \hline
$2 \boldsymbol{\gamma}$ & \multicolumn{1}{c|}{$(0,0,0,0)$}                                                                                           & \multicolumn{1}{c|}{$~~~(1,1,1,1)~~~$}                                 & \multicolumn{1}{c|}{\begin{tabular}[c]{@{}c@{}}$~~~~(1,0,1,0)~~~~$\\ $~~~(0,1,0,1)~~~$\end{tabular}}                                                                  & \multicolumn{1}{c|}{$~(\alpha,1,\bar{\alpha},\beta)^*~$} & $~(\alpha,\alpha,\beta,\bar{\beta})^\dagger$ \\ \hline
Pattern                                       & \multicolumn{1}{c|}{\begin{tabular}[c]{@{}c@{}}Ammann-Beenker\\ Fig.\ref{fig: window} (c)\end{tabular}} & \multicolumn{1}{c|}{Fig.\ref{fig: summary} (a)} & \multicolumn{1}{c|}{\begin{tabular}[c]{@{}c@{}}Fig.\ref{fig: summary} (b)\\ (Fig.\ref{fig: window} (b))\end{tabular}} & \multicolumn{1}{c|}{Fig.\ref{fig: summary} (c)}                                                                                       & Fig.\ref{fig: summary} (d)    \\           
\hline                             
\end{tabular}
\caption{\label{talbe: summary} Classification of the commensurate shift vector $\boldsymbol{\gamma}$ in hypercubic lattice and the relevant new quasicrystalline patterns. Depending on $\boldsymbol{\gamma}$, new quasicrystals with 2-, 4- and 8-fold rotational symmetries emerge as shown in Figs.\ref{fig: summary}(a)-(d). $*$ and $\dagger$ represent the sets including: $(~~)^*=\{(\beta,\alpha,1,\bar{\alpha}), (1,\alpha,\beta,\alpha), (\alpha,\beta,\alpha,1)\}$ and $(~~)^\dagger=\{ (\alpha,\beta,\beta,\alpha), (\alpha,\bar{\alpha},\beta,\beta), (\alpha,\beta,\bar{\beta},\bar{\alpha})\}$, where $\bar{a}\equiv-a  \mbox{ mod } 1$ and $\alpha,\beta$ are arbitrary rational numbers.}
\label{table: summary}
\end{table*}

\section{Discussions and conclusion}
\label{sec: discussion}
In summary, new quasi-periodic patterns with distinct rotational symmetries are discovered via translation operations of the high-dimensional lattice, as illustrated in Fig.\ref{fig: window}.
Focusing on the four-dimensional hypercubic lattice and its special translations, one finds completely new quasi-periodic patterns in the projected two-dimensional space, where different global rotational symmetries emerge,  as summarized in Fig.\ref{fig: summary} and Table.\ref{table: summary}.  
It offers a systematic approach to discover new types of quasicrystals via a simple operation in the high-dimensional lattice and to control symmetries in quasi-periodic systems in addition to their classifications.

Experimental realization of these new quasi-periodic patterns can be also examined in various systems such as the optical lattice and the Floquet lattice with multiple frequency driving systems\cite{zhao2013hamiltonian,jagannathan2014eightfold,corcovilos2019two,sanchez2005bose,viebahn2019matter,sbroscia2020observing,russomanno2017floquet,martin2017topological,rodriguez2021real,malz2021topological,poertner2020validity,thomson2021flow}. 
Particularly, with control of fine-tuned multiple driving frequencies, the pseudo-electric field in the Floquet system enables  the formation of quasicrystals under the projection of high-dimension. Furthermore, tuning frequencies also allows translation operation in the Floquet system thus resulting in new quasi-periodic structures. \cite{martin2017topological,malz2021topological,poertner2020validity,thomson2021flow}. 
In addition, our study is also applicable to other higher-order symmetric quasi-periodic systems such as decagonal or dodecagonal quasicrystals in two-dimension, icosahedral quasicrystals in three-dimension and more general quasicrystals originated from the root lattices of the Lie groups\cite{tracey2021programming,van2012formation,baake2017aperiodic,sadoc1993e8,yu2019dodecagonal}. 
With the realization of new quasicrystals, one can further demonstrate physical phenomena originated from the high-dimensional degree of freedom. Thus, the characteristic symmetries of quasicrystalline patterns could be controlled via special operations of high-dimensional lattice. Along with the discovery of new quasicrystals, one could also find exotic quantum phenomenon that is uniquely present in those quasicrystals and we leave it as an interesting future work\footnote{Junmo Jeon and SungBin Lee, In preparation}.




\section{Methods}
\label{sec:method}
To demonstrate new quasicrystals, the four-dimensional hypercubic lattice is constructed whose unit length is 1. The size of the hypercubic lattice is 50, and hence there are $50^4$ number of hypercubic cells. Specifically, each axis runs from $-25$ to 25. By using the projection maps, Eq.\eqref{projection} and \eqref{projectionperp}, we project hypercubic lattice points onto the two-dimensional surfaces, physical space $\boldsymbol{\pi}$ and perpendicular space $\boldsymbol{\pi}^\perp$, respectively. The hypercubic lattice points are selected whose the projection image on the perpendicular space belongs to the window $K$, which is the projection of the Wigner-Seitz cell of the origin. The tiling pattern is formed by connecting all two points that are separated by a distance of 1 with a line. We repeat this process with different translations of the hypercubic lattice and get the distinct quasicrystalline tiling patterns and symmetry.

\section*{Acknowledgements}
This work is supported by National Research Foundation Grant (NRF-2020R1A4A3079707, NRF grand 2021R1A2C1093060).

\clearpage
\pagebreak

\renewcommand{\thesection}{\arabic{section}}
\setcounter{section}{0}
\renewcommand{\thefigure}{S\arabic{figure}}
\setcounter{figure}{0}
\renewcommand{\theequation}{S\arabic{equation}}
\setcounter{equation}{0}

\section*{Supplementary Material}

\maketitle

\section{$\pi^{\perp}$-projection images}
\label{sec: 1}

\begin{figure*}[]
\centering
\includegraphics[width=0.9\textwidth]{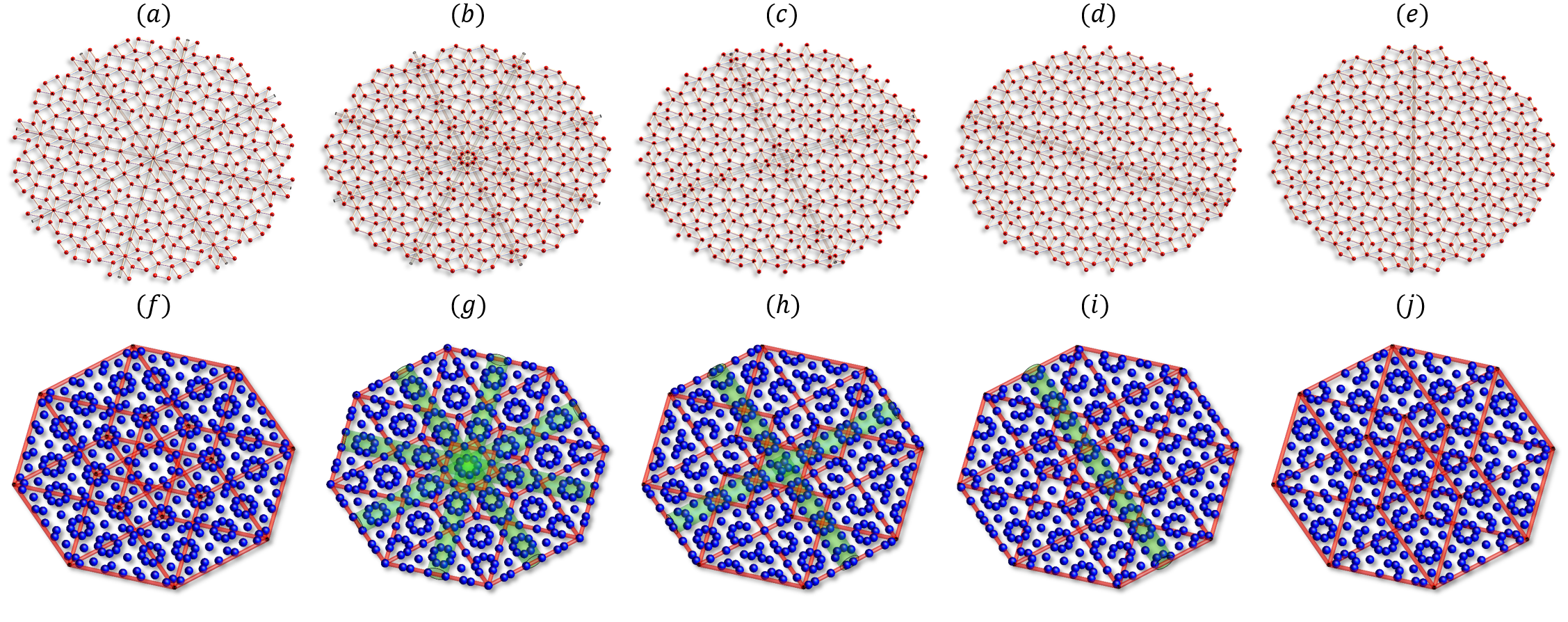}
\label{fig1}
\caption{\label{fig: windows} Depending on translation operation in the four-dimensional hypercubic lattice, (a)-(e) various types of quasicrystals on the physical space ($\boldsymbol{\pi}$-projection) and (f)-(j)  their perpendicular space ($\boldsymbol{\pi}^\perp$-projection) images.
Here, the translation vector $\boldsymbol{\gamma}$ is given by (a),(f) $\boldsymbol{\gamma}=\boldsymbol{0}$ (b),(g) $\boldsymbol{\gamma}=\frac{1}{2}(1,1,1,1)$ (c),(h) $\boldsymbol{\gamma}=\frac{1}{2}(1,0,1,0)$ (d),(i) $\boldsymbol{\gamma}=\frac{1}{2}(1,0,0,0)$, and (e),(j) $\boldsymbol{\gamma}=\frac{1}{2}(1,1,0,0)$, respectively. The symmetries of both $\boldsymbol{\pi}$-, and $\boldsymbol{\pi}^{\perp}$-projection images are the same as (a),(b),(f),(g) 8-fold, (c),(h) 4-fold, and (d),(e),(i),(j) 2-fold, respectively. The gray lines in (a)-(e) are drawn for guide to the eyes, to emphasize the rotational symmetries. The red lines in (f)-(j) are drawn for the window formed by the $\boldsymbol{\pi}^\perp$-projection of the Wigner-Seitz cell. The green shaded regions in (g)-(i) correspond to the new local pattern along the gray lines in the quasicrystals illustrated in (b)-(d), respectively. See the main text for detailed information.
}
\end{figure*}

In this section, we discuss the perpendicular space $\boldsymbol{\pi}^\perp$ projection images for each quasicrystal discussed in the main text. 
Figure \ref{fig: windows} displays each quasicrystal structure ((a)-(e)) and their $\boldsymbol{\pi}^{\perp}$-projection images on the window ((f)-(j)). Figure \ref{fig: windows} (a) is the Ammann-Beenker tiling. Figures \ref{fig: windows} (b)-(e) are new types of quasicrystal structure due to the translations of the $4$-dimensional hypercubic lattice with the shift vectors, $\boldsymbol{\gamma}$ : (b) $\boldsymbol{\gamma}= \frac{1}{2}(1,1,1,1)$ (c) $\boldsymbol{\gamma}=\frac{1}{2}(1,0,1,0)$ (d) $\boldsymbol{\gamma}=\frac{1}{2}(1,0,0,0)$ and (e) $\boldsymbol{\gamma}=\frac{1}{2}(1,1,0,0)$, respectively. Here, the gray lines in Figures \ref{fig: windows} (a)-(e) are drawn for guide to the eyes, to emphasize their rotational symmetries. Figures \ref{fig: windows} (f)-(j) are the $\boldsymbol{\pi}^\perp$-projection images for Figures \ref{fig: windows} (a)-(e), respectively. Note that the $\boldsymbol{\pi}^{\perp}$-projection images share the same symmetry with the quasicrystals. Thus, there are only 2-,4- and 8-fold rotational symmetric quasicrystals since $\boldsymbol{\pi}^\perp$-projection images densely cover the window in thermodynamic limit.
Furthermore, $\boldsymbol{\pi}^{\perp}$-projection also specifies the new local patterns along the diagonals in the quasicrystals as shown in Figures \ref{fig: windows} (b)-(d). To be more specific, the new local pattern that is present along the gray lines in Figures \ref{fig: windows} (b)-(d), correspond to the $\boldsymbol{\pi}^{\perp}$-projection images on the green shaded regions in Figures \ref{fig: windows} (g)-(i), respectively. On the other hand, the $\boldsymbol{\pi}^{\perp}$ images for both Figures \ref{fig: windows}(f) and (j) do not exhibit such green shaded region. This is relevant for the absence of new local pattern in the Ammann-Beenker tiling (Figure \ref{fig: windows} (a)) and the quasicrystal shown in Figure \ref{fig: windows} (j).  

\section{Asymmetric quasicrystal originated from the arbitrary commensurate translation}
\label{sec: 2}

\begin{figure*}[]
\centering
\includegraphics[width=0.9\textwidth]{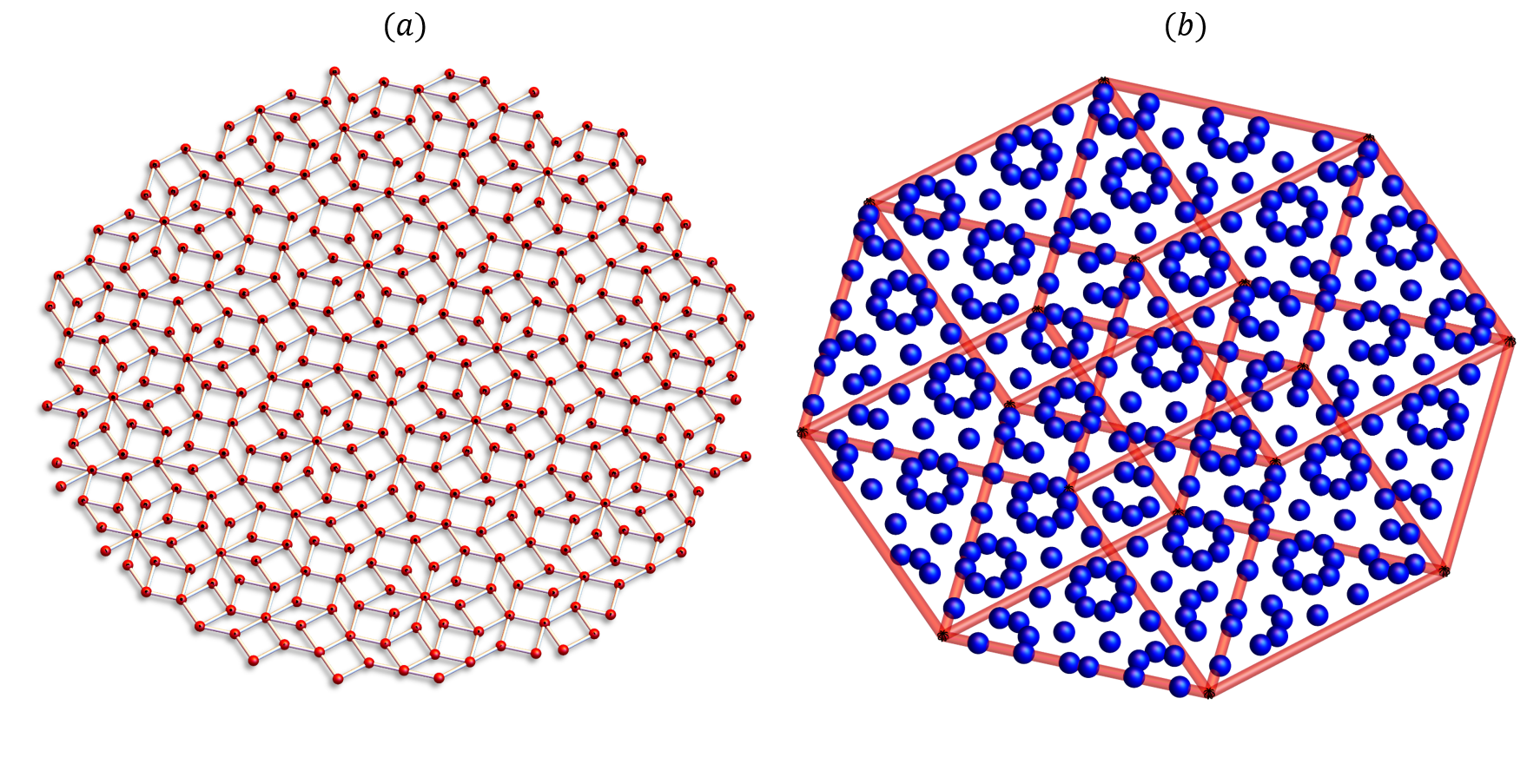}
\label{fig1}
\caption{\label{fig: asymmetric} 
For a specific translation  
$\boldsymbol{\gamma}=(\frac{1}{2},\frac{1}{3},\frac{1}{4},\frac{1}{5})$ in the hypercubic lattice, 
(a) Asymmetric tiling pattern on the physical space image ($\boldsymbol{\pi}$-projection) and (b) the perpendicular space image on the window ($\boldsymbol{\pi}^\perp$-projection). The red lines are the $\boldsymbol{\pi}^\perp$-projection of the Wigner-Seitz cell.}
\end{figure*}

For comparison with the special sets of $\boldsymbol{\gamma}$ described above and in the main text, Figure \ref{fig: asymmetric} illustrates an example of the projection images for the arbitrary commensurate translation with a shift vector $\boldsymbol{\gamma}$ as $(\frac{1}{2},\frac{1}{3},\frac{1}{4},\frac{1}{5})$ : (a) the quasicrystal ($\boldsymbol{\pi}$-projection) and (b) it's perpendicular space image on the window ($\boldsymbol{\pi}^\perp$-projection). 
 Although Figure \ref{fig: asymmetric} possesses local 8-fold symmetric patterns, the global rotational symmetry is absent and it becomes asymmetric. Note that the $\boldsymbol{\pi}^{\perp}$-projection image on the window in Figure \ref{fig: asymmetric} (b) is also asymmetric. In addition, it does not possess any new local pattern shown in Figures \ref{fig: windows} (b)-(d). The asymmetric projection images exemplified in Figure \ref{fig: asymmetric} are general for the arbitrary commensurate translations not classified by Table. 1 in the main text.



\bibliography{reference}

\begin{thebibliography}{71}%
\makeatletter
\providecommand \@ifxundefined [1]{%
 \@ifx{#1\undefined}
}%
\providecommand \@ifnum [1]{%
 \ifnum #1\expandafter \@firstoftwo
 \else \expandafter \@secondoftwo
 \fi
}%
\providecommand \@ifx [1]{%
 \ifx #1\expandafter \@firstoftwo
 \else \expandafter \@secondoftwo
 \fi
}%
\providecommand \natexlab [1]{#1}%
\providecommand \enquote  [1]{``#1''}%
\providecommand \bibnamefont  [1]{#1}%
\providecommand \bibfnamefont [1]{#1}%
\providecommand \citenamefont [1]{#1}%
\providecommand \href@noop [0]{\@secondoftwo}%
\providecommand \href [0]{\begingroup \@sanitize@url \@href}%
\providecommand \@href[1]{\@@startlink{#1}\@@href}%
\providecommand \@@href[1]{\endgroup#1\@@endlink}%
\providecommand \@sanitize@url [0]{\catcode `\\12\catcode `\$12\catcode
  `\&12\catcode `\#12\catcode `\^12\catcode `\_12\catcode `\%12\relax}%
\providecommand \@@startlink[1]{}%
\providecommand \@@endlink[0]{}%
\providecommand \url  [0]{\begingroup\@sanitize@url \@url }%
\providecommand \@url [1]{\endgroup\@href {#1}{\urlprefix }}%
\providecommand \urlprefix  [0]{URL }%
\providecommand \Eprint [0]{\href }%
\providecommand \doibase [0]{http://dx.doi.org/}%
\providecommand \selectlanguage [0]{\@gobble}%
\providecommand \bibinfo  [0]{\@secondoftwo}%
\providecommand \bibfield  [0]{\@secondoftwo}%
\providecommand \translation [1]{[#1]}%
\providecommand \BibitemOpen [0]{}%
\providecommand \bibitemStop [0]{}%
\providecommand \bibitemNoStop [0]{.\EOS\space}%
\providecommand \EOS [0]{\spacefactor3000\relax}%
\providecommand \BibitemShut  [1]{\csname bibitem#1\endcsname}%
\let\auto@bib@innerbib\@empty
\bibitem [{\citenamefont {Chaikin}\ \emph {et~al.}(1995)\citenamefont
  {Chaikin}, \citenamefont {Lubensky},\ and\ \citenamefont
  {Witten}}]{chaikin1995principles}%
  \BibitemOpen
  \bibfield  {author} {\bibinfo {author} {\bibfnamefont {Paul~M}\ \bibnamefont
  {Chaikin}}, \bibinfo {author} {\bibfnamefont {Tom~C}\ \bibnamefont
  {Lubensky}}, \ and\ \bibinfo {author} {\bibfnamefont {Thomas~A}\ \bibnamefont
  {Witten}},\ }\href@noop {} {\emph {\bibinfo {title} {Principles of condensed
  matter physics}}},\ Vol.~\bibinfo {volume} {10}\ (\bibinfo  {publisher}
  {Cambridge university press Cambridge},\ \bibinfo {year} {1995})\BibitemShut
  {NoStop}%
\bibitem [{\citenamefont {Tsvelik}(2007)}]{tsvelik2007quantum}%
  \BibitemOpen
  \bibfield  {author} {\bibinfo {author} {\bibfnamefont {Alexei~M}\
  \bibnamefont {Tsvelik}},\ }\href@noop {} {\emph {\bibinfo {title} {Quantum
  field theory in condensed matter physics}}}\ (\bibinfo  {publisher}
  {Cambridge university press},\ \bibinfo {year} {2007})\BibitemShut {NoStop}%
\bibitem [{\citenamefont {Marder}(2010)}]{marder2010condensed}%
  \BibitemOpen
  \bibfield  {author} {\bibinfo {author} {\bibfnamefont {Michael~P}\
  \bibnamefont {Marder}},\ }\href@noop {} {\emph {\bibinfo {title} {Condensed
  matter physics}}}\ (\bibinfo  {publisher} {John Wiley \& Sons},\ \bibinfo
  {year} {2010})\BibitemShut {NoStop}%
\bibitem [{\citenamefont {Simon}(2013)}]{simon2013oxford}%
  \BibitemOpen
  \bibfield  {author} {\bibinfo {author} {\bibfnamefont {Steven~H}\
  \bibnamefont {Simon}},\ }\href@noop {} {\emph {\bibinfo {title} {The Oxford
  solid state basics}}}\ (\bibinfo  {publisher} {OUP Oxford},\ \bibinfo {year}
  {2013})\BibitemShut {NoStop}%
\bibitem [{\citenamefont {Jones}\ and\ \citenamefont
  {March}(1985)}]{jones1985theoretical}%
  \BibitemOpen
  \bibfield  {author} {\bibinfo {author} {\bibfnamefont {William}\ \bibnamefont
  {Jones}}\ and\ \bibinfo {author} {\bibfnamefont {Norman~Henry}\ \bibnamefont
  {March}},\ }\href@noop {} {\emph {\bibinfo {title} {Theoretical solid state
  physics}}},\ Vol.~\bibinfo {volume} {35}\ (\bibinfo  {publisher} {Courier
  Corporation},\ \bibinfo {year} {1985})\BibitemShut {NoStop}%
\bibitem [{\citenamefont {Cohen}\ and\ \citenamefont
  {Louie}(2016)}]{cohen2016fundamentals}%
  \BibitemOpen
  \bibfield  {author} {\bibinfo {author} {\bibfnamefont {Marvin~L}\
  \bibnamefont {Cohen}}\ and\ \bibinfo {author} {\bibfnamefont {Steven~G}\
  \bibnamefont {Louie}},\ }\href@noop {} {\emph {\bibinfo {title} {Fundamentals
  of condensed matter physics}}}\ (\bibinfo  {publisher} {Cambridge University
  Press},\ \bibinfo {year} {2016})\BibitemShut {NoStop}%
\bibitem [{\citenamefont {De~Graef}\ and\ \citenamefont
  {McHenry}(2012)}]{de2012structure}%
  \BibitemOpen
  \bibfield  {author} {\bibinfo {author} {\bibfnamefont {Marc}\ \bibnamefont
  {De~Graef}}\ and\ \bibinfo {author} {\bibfnamefont {Michael~E}\ \bibnamefont
  {McHenry}},\ }\href@noop {} {\emph {\bibinfo {title} {Structure of materials:
  an introduction to crystallography, diffraction and symmetry}}}\ (\bibinfo
  {publisher} {Cambridge University Press},\ \bibinfo {year}
  {2012})\BibitemShut {NoStop}%
\bibitem [{\citenamefont {Haldane}(1985)}]{haldane1985many}%
  \BibitemOpen
  \bibfield  {author} {\bibinfo {author} {\bibfnamefont {FDM}\ \bibnamefont
  {Haldane}},\ }\bibfield  {title} {\enquote {\bibinfo {title} {Many-particle
  translational symmetries of two-dimensional electrons at rational
  landau-level filling},}\ }\href@noop {} {\bibfield  {journal} {\bibinfo
  {journal} {Physical Review Letters}\ }\textbf {\bibinfo {volume} {55}},\
  \bibinfo {pages} {2095} (\bibinfo {year} {1985})}\BibitemShut {NoStop}%
\bibitem [{\citenamefont {Bernard}\ and\ \citenamefont
  {Felder}(1991)}]{bernard1991quantum}%
  \BibitemOpen
  \bibfield  {author} {\bibinfo {author} {\bibfnamefont {Denis}\ \bibnamefont
  {Bernard}}\ and\ \bibinfo {author} {\bibfnamefont {Giovanni}\ \bibnamefont
  {Felder}},\ }\bibfield  {title} {\enquote {\bibinfo {title} {Quantum group
  symmetries in two-dimensional lattice quantum field theory},}\ }\href@noop {}
  {\bibfield  {journal} {\bibinfo  {journal} {Nuclear Physics B, Particle
  Physics}\ }\textbf {\bibinfo {volume} {365}},\ \bibinfo {pages} {98--120}
  (\bibinfo {year} {1991})}\BibitemShut {NoStop}%
\bibitem [{\citenamefont {Kittel}\ and\ \citenamefont
  {McEuen}(1976)}]{kittel1976introduction}%
  \BibitemOpen
  \bibfield  {author} {\bibinfo {author} {\bibfnamefont {Charles}\ \bibnamefont
  {Kittel}}\ and\ \bibinfo {author} {\bibfnamefont {Paul}\ \bibnamefont
  {McEuen}},\ }\href@noop {} {\emph {\bibinfo {title} {Introduction to solid
  state physics}}},\ Vol.~\bibinfo {volume} {8}\ (\bibinfo  {publisher} {Wiley
  New York},\ \bibinfo {year} {1976})\BibitemShut {NoStop}%
\bibitem [{\citenamefont {Estrada}\ \emph {et~al.}(2009)\citenamefont
  {Estrada}, \citenamefont {Candelas}, \citenamefont {Uris}, \citenamefont
  {Belmar}, \citenamefont {Garc{\'\i}a~de Abajo},\ and\ \citenamefont
  {Meseguer}}]{estrada2009influence}%
  \BibitemOpen
  \bibfield  {author} {\bibinfo {author} {\bibfnamefont {H{\'e}ctor}\
  \bibnamefont {Estrada}}, \bibinfo {author} {\bibfnamefont {Pilar}\
  \bibnamefont {Candelas}}, \bibinfo {author} {\bibfnamefont {Antonio}\
  \bibnamefont {Uris}}, \bibinfo {author} {\bibfnamefont {Francisco}\
  \bibnamefont {Belmar}}, \bibinfo {author} {\bibfnamefont {F~Javier}\
  \bibnamefont {Garc{\'\i}a~de Abajo}}, \ and\ \bibinfo {author} {\bibfnamefont
  {Francisco}\ \bibnamefont {Meseguer}},\ }\bibfield  {title} {\enquote
  {\bibinfo {title} {Influence of lattice symmetry on ultrasound transmission
  through plates with subwavelength aperture arrays},}\ }\href@noop {}
  {\bibfield  {journal} {\bibinfo  {journal} {Applied Physics Letters}\
  }\textbf {\bibinfo {volume} {95}},\ \bibinfo {pages} {051906} (\bibinfo
  {year} {2009})}\BibitemShut {NoStop}%
\bibitem [{\citenamefont {Petsas}\ \emph {et~al.}(1994)\citenamefont {Petsas},
  \citenamefont {Coates},\ and\ \citenamefont
  {Grynberg}}]{petsas1994crystallography}%
  \BibitemOpen
  \bibfield  {author} {\bibinfo {author} {\bibfnamefont {KI}~\bibnamefont
  {Petsas}}, \bibinfo {author} {\bibfnamefont {AB}~\bibnamefont {Coates}}, \
  and\ \bibinfo {author} {\bibfnamefont {G}~\bibnamefont {Grynberg}},\
  }\bibfield  {title} {\enquote {\bibinfo {title} {Crystallography of optical
  lattices},}\ }\href@noop {} {\bibfield  {journal} {\bibinfo  {journal}
  {Physical review A}\ }\textbf {\bibinfo {volume} {50}},\ \bibinfo {pages}
  {5173} (\bibinfo {year} {1994})}\BibitemShut {NoStop}%
\bibitem [{\citenamefont {MacChesney}\ \emph {et~al.}(1965)\citenamefont
  {MacChesney}, \citenamefont {Sherwood},\ and\ \citenamefont
  {Potter}}]{macchesney1965electric}%
  \BibitemOpen
  \bibfield  {author} {\bibinfo {author} {\bibfnamefont {JB}~\bibnamefont
  {MacChesney}}, \bibinfo {author} {\bibfnamefont {RC}~\bibnamefont
  {Sherwood}}, \ and\ \bibinfo {author} {\bibfnamefont {JF}~\bibnamefont
  {Potter}},\ }\bibfield  {title} {\enquote {\bibinfo {title} {Electric and
  magnetic properties of the strontium ferrates},}\ }\href@noop {} {\bibfield
  {journal} {\bibinfo  {journal} {The Journal of Chemical Physics}\ }\textbf
  {\bibinfo {volume} {43}},\ \bibinfo {pages} {1907--1913} (\bibinfo {year}
  {1965})}\BibitemShut {NoStop}%
\bibitem [{\citenamefont {Dmitriev}(2005)}]{dmitriev2005symmetry}%
  \BibitemOpen
  \bibfield  {author} {\bibinfo {author} {\bibfnamefont {V}~\bibnamefont
  {Dmitriev}},\ }\bibfield  {title} {\enquote {\bibinfo {title} {Symmetry
  properties of 2d magnetic photonic crystals with square lattice},}\
  }\href@noop {} {\bibfield  {journal} {\bibinfo  {journal} {The European
  Physical Journal-Applied Physics}\ }\textbf {\bibinfo {volume} {32}},\
  \bibinfo {pages} {159--165} (\bibinfo {year} {2005})}\BibitemShut {NoStop}%
\bibitem [{\citenamefont {Khan}\ \emph {et~al.}(2019)\citenamefont {Khan},
  \citenamefont {Prishchenko}, \citenamefont {Skourski}, \citenamefont
  {Mazurenko},\ and\ \citenamefont {Tsirlin}}]{khan2019cubic}%
  \BibitemOpen
  \bibfield  {author} {\bibinfo {author} {\bibfnamefont {Nazir}\ \bibnamefont
  {Khan}}, \bibinfo {author} {\bibfnamefont {Danil}\ \bibnamefont
  {Prishchenko}}, \bibinfo {author} {\bibfnamefont {Yurii}\ \bibnamefont
  {Skourski}}, \bibinfo {author} {\bibfnamefont {Vladimir~G}\ \bibnamefont
  {Mazurenko}}, \ and\ \bibinfo {author} {\bibfnamefont {Alexander~A}\
  \bibnamefont {Tsirlin}},\ }\bibfield  {title} {\enquote {\bibinfo {title}
  {Cubic symmetry and magnetic frustration on the fcc spin lattice in k 2 ircl
  6},}\ }\href@noop {} {\bibfield  {journal} {\bibinfo  {journal} {Physical
  Review B}\ }\textbf {\bibinfo {volume} {99}},\ \bibinfo {pages} {144425}
  (\bibinfo {year} {2019})}\BibitemShut {NoStop}%
\bibitem [{\citenamefont {Yamamoto}(1996)}]{yamamoto1996crystallography}%
  \BibitemOpen
  \bibfield  {author} {\bibinfo {author} {\bibfnamefont {Akiji}\ \bibnamefont
  {Yamamoto}},\ }\bibfield  {title} {\enquote {\bibinfo {title}
  {Crystallography of quasiperiodic crystals},}\ }\href@noop {} {\bibfield
  {journal} {\bibinfo  {journal} {Acta Crystallographica Section A: Foundations
  of Crystallography}\ }\textbf {\bibinfo {volume} {52}},\ \bibinfo {pages}
  {509--560} (\bibinfo {year} {1996})}\BibitemShut {NoStop}%
\bibitem [{\citenamefont {Baake}\ and\ \citenamefont
  {Grimm}(2017)}]{baake2017aperiodic}%
  \BibitemOpen
  \bibfield  {author} {\bibinfo {author} {\bibfnamefont {Michael}\ \bibnamefont
  {Baake}}\ and\ \bibinfo {author} {\bibfnamefont {Uwe}\ \bibnamefont
  {Grimm}},\ }\href@noop {} {\emph {\bibinfo {title} {Aperiodic Order: Volume
  2, Crystallography and Almost Periodicity}}},\ Vol.\ \bibinfo {volume} {166}\
  (\bibinfo  {publisher} {Cambridge University Press},\ \bibinfo {year}
  {2017})\BibitemShut {NoStop}%
\bibitem [{\citenamefont {Steurer}\ and\ \citenamefont
  {Deloudi}(2009)}]{walter2009crystallography}%
  \BibitemOpen
  \bibfield  {author} {\bibinfo {author} {\bibfnamefont {Walter}\ \bibnamefont
  {Steurer}}\ and\ \bibinfo {author} {\bibfnamefont {Sofia}\ \bibnamefont
  {Deloudi}},\ }\href@noop {} {\emph {\bibinfo {title} {Crystallography of
  quasicrystals: concepts, methods and structures}}},\ Vol.\ \bibinfo {volume}
  {126}\ (\bibinfo  {publisher} {Springer Science \& Business Media},\ \bibinfo
  {year} {2009})\BibitemShut {NoStop}%
\bibitem [{\citenamefont {Fuentes}\ \emph {et~al.}(2004)\citenamefont
  {Fuentes}, \citenamefont {Garc{\i}a}, \citenamefont {Matutes-Aquino},\ and\
  \citenamefont {R{\i}os-Jara}}]{fuentes2004magnetoelectricity}%
  \BibitemOpen
  \bibfield  {author} {\bibinfo {author} {\bibfnamefont {L}~\bibnamefont
  {Fuentes}}, \bibinfo {author} {\bibfnamefont {M}~\bibnamefont {Garc{\i}a}},
  \bibinfo {author} {\bibfnamefont {J}~\bibnamefont {Matutes-Aquino}}, \ and\
  \bibinfo {author} {\bibfnamefont {D}~\bibnamefont {R{\i}os-Jara}},\
  }\bibfield  {title} {\enquote {\bibinfo {title} {Magnetoelectricity via
  crystallography},}\ }\href@noop {} {\bibfield  {journal} {\bibinfo  {journal}
  {Journal of alloys and compounds}\ }\textbf {\bibinfo {volume} {369}},\
  \bibinfo {pages} {10--13} (\bibinfo {year} {2004})}\BibitemShut {NoStop}%
\bibitem [{\citenamefont {Das}\ \emph {et~al.}(2020)\citenamefont {Das},
  \citenamefont {Alam},\ and\ \citenamefont
  {Akther}}]{das2020crystallographic}%
  \BibitemOpen
  \bibfield  {author} {\bibinfo {author} {\bibfnamefont {Bablu~Chandra}\
  \bibnamefont {Das}}, \bibinfo {author} {\bibfnamefont {F}~\bibnamefont
  {Alam}}, \ and\ \bibinfo {author} {\bibfnamefont {AKM~Hossain}\ \bibnamefont
  {Akther}},\ }\bibfield  {title} {\enquote {\bibinfo {title} {The
  crystallographic, magnetic, and electrical properties of gd3+-substituted
  ni--cu--zn mixed ferrites},}\ }\href@noop {} {\bibfield  {journal} {\bibinfo
  {journal} {Journal of Physics and Chemistry of Solids}\ }\textbf {\bibinfo
  {volume} {142}},\ \bibinfo {pages} {109433} (\bibinfo {year}
  {2020})}\BibitemShut {NoStop}%
\bibitem [{\citenamefont {Routledge}\ \emph {et~al.}(2021)\citenamefont
  {Routledge}, \citenamefont {Vir}, \citenamefont {Cook}, \citenamefont
  {Murgatroyd}, \citenamefont {Ahmed}, \citenamefont {Savvin}, \citenamefont
  {Claridge},\ and\ \citenamefont {Alaria}}]{routledge2021mode}%
  \BibitemOpen
  \bibfield  {author} {\bibinfo {author} {\bibfnamefont {Kieran}\ \bibnamefont
  {Routledge}}, \bibinfo {author} {\bibfnamefont {Praveen}\ \bibnamefont
  {Vir}}, \bibinfo {author} {\bibfnamefont {Nicholas}\ \bibnamefont {Cook}},
  \bibinfo {author} {\bibfnamefont {Philip~AE}\ \bibnamefont {Murgatroyd}},
  \bibinfo {author} {\bibfnamefont {Sheikh~J}\ \bibnamefont {Ahmed}}, \bibinfo
  {author} {\bibfnamefont {Stanislav~N}\ \bibnamefont {Savvin}}, \bibinfo
  {author} {\bibfnamefont {John~B}\ \bibnamefont {Claridge}}, \ and\ \bibinfo
  {author} {\bibfnamefont {Jonathan}\ \bibnamefont {Alaria}},\ }\bibfield
  {title} {\enquote {\bibinfo {title} {Mode crystallography analysis through
  the structural phase transition and magnetic critical behavior of the lacunar
  spinel gamo4se8},}\ }\href@noop {} {\bibfield  {journal} {\bibinfo  {journal}
  {Chemistry of Materials}\ }\textbf {\bibinfo {volume} {33}},\ \bibinfo
  {pages} {5718--5729} (\bibinfo {year} {2021})}\BibitemShut {NoStop}%
\bibitem [{\citenamefont {Perez-Mato}\ \emph {et~al.}(2015)\citenamefont
  {Perez-Mato}, \citenamefont {Gallego}, \citenamefont {Tasci}, \citenamefont
  {Elcoro}, \citenamefont {de~la Flor},\ and\ \citenamefont
  {Aroyo}}]{perez2015symmetry}%
  \BibitemOpen
  \bibfield  {author} {\bibinfo {author} {\bibfnamefont {JM}~\bibnamefont
  {Perez-Mato}}, \bibinfo {author} {\bibfnamefont {SV}~\bibnamefont {Gallego}},
  \bibinfo {author} {\bibfnamefont {ES}~\bibnamefont {Tasci}}, \bibinfo
  {author} {\bibfnamefont {LU{\.I}S}\ \bibnamefont {Elcoro}}, \bibinfo {author}
  {\bibfnamefont {Gemma}\ \bibnamefont {de~la Flor}}, \ and\ \bibinfo {author}
  {\bibfnamefont {MI}~\bibnamefont {Aroyo}},\ }\bibfield  {title} {\enquote
  {\bibinfo {title} {Symmetry-based computational tools for magnetic
  crystallography},}\ }\href@noop {} {\bibfield  {journal} {\bibinfo  {journal}
  {Annual Review of Materials Research}\ }\textbf {\bibinfo {volume} {45}},\
  \bibinfo {pages} {217--248} (\bibinfo {year} {2015})}\BibitemShut {NoStop}%
\bibitem [{\citenamefont {Wang}\ \emph {et~al.}(1987)\citenamefont {Wang},
  \citenamefont {Chen},\ and\ \citenamefont {Kuo}}]{PhysRevLett.59.1010}%
  \BibitemOpen
  \bibfield  {author} {\bibinfo {author} {\bibfnamefont {N.}~\bibnamefont
  {Wang}}, \bibinfo {author} {\bibfnamefont {H.}~\bibnamefont {Chen}}, \ and\
  \bibinfo {author} {\bibfnamefont {K.~H.}\ \bibnamefont {Kuo}},\ }\bibfield
  {title} {\enquote {\bibinfo {title} {Two-dimensional quasicrystal with
  eightfold rotational symmetry},}\ }\href@noop {} {\bibfield  {journal}
  {\bibinfo  {journal} {Phys. Rev. Lett.}\ }\textbf {\bibinfo {volume} {59}},\
  \bibinfo {pages} {1010--1013} (\bibinfo {year} {1987})}\BibitemShut {NoStop}%
\bibitem [{\citenamefont {Bursill}\ and\ \citenamefont
  {Lin}(1985)}]{bursill1985penrose}%
  \BibitemOpen
  \bibfield  {author} {\bibinfo {author} {\bibfnamefont {LA}~\bibnamefont
  {Bursill}}\ and\ \bibinfo {author} {\bibfnamefont {Peng~Ju}\ \bibnamefont
  {Lin}},\ }\bibfield  {title} {\enquote {\bibinfo {title} {Penrose tiling
  observed in a quasi-crystal},}\ }\href@noop {} {\bibfield  {journal}
  {\bibinfo  {journal} {Nature}\ }\textbf {\bibinfo {volume} {316}},\ \bibinfo
  {pages} {50--51} (\bibinfo {year} {1985})}\BibitemShut {NoStop}%
\bibitem [{\citenamefont {Reinhardt}\ \emph {et~al.}(2016)\citenamefont
  {Reinhardt}, \citenamefont {Schreck}, \citenamefont {Romano},\ and\
  \citenamefont {Doye}}]{reinhardt2016self}%
  \BibitemOpen
  \bibfield  {author} {\bibinfo {author} {\bibfnamefont {Aleks}\ \bibnamefont
  {Reinhardt}}, \bibinfo {author} {\bibfnamefont {John~S}\ \bibnamefont
  {Schreck}}, \bibinfo {author} {\bibfnamefont {Flavio}\ \bibnamefont
  {Romano}}, \ and\ \bibinfo {author} {\bibfnamefont {Jonathan~PK}\
  \bibnamefont {Doye}},\ }\bibfield  {title} {\enquote {\bibinfo {title}
  {Self-assembly of two-dimensional binary quasicrystals: A possible route to a
  dna quasicrystal},}\ }\href@noop {} {\bibfield  {journal} {\bibinfo
  {journal} {Journal of Physics: Condensed Matter}\ }\textbf {\bibinfo {volume}
  {29}},\ \bibinfo {pages} {014006} (\bibinfo {year} {2016})}\BibitemShut
  {NoStop}%
\bibitem [{\citenamefont {Grimm}\ and\ \citenamefont
  {Scheffer}(2003)}]{GRIMM2003731}%
  \BibitemOpen
  \bibfield  {author} {\bibinfo {author} {\bibfnamefont {Uwe}\ \bibnamefont
  {Grimm}}\ and\ \bibinfo {author} {\bibfnamefont {Max}\ \bibnamefont
  {Scheffer}},\ }\bibfield  {title} {\enquote {\bibinfo {title} {Incommensurate
  crystals and quasicrystals},}\ }in\ \href@noop {} {\emph {\bibinfo
  {booktitle} {Encyclopedia of Physical Science and Technology (Third
  Edition)}}},\ \bibinfo {editor} {edited by\ \bibinfo {editor} {\bibfnamefont
  {Robert~A.}\ \bibnamefont {Meyers}}}\ (\bibinfo  {publisher} {Academic
  Press},\ \bibinfo {address} {New York},\ \bibinfo {year} {2003})\ \bibinfo
  {edition} {third edition}\ ed.,\ pp.\ \bibinfo {pages} {731--749}\BibitemShut
  {NoStop}%
\bibitem [{\citenamefont {Noya}\ \emph {et~al.}(2021)\citenamefont {Noya},
  \citenamefont {Wong}, \citenamefont {Llombart},\ and\ \citenamefont
  {Doye}}]{noya2021design}%
  \BibitemOpen
  \bibfield  {author} {\bibinfo {author} {\bibfnamefont {Eva~G}\ \bibnamefont
  {Noya}}, \bibinfo {author} {\bibfnamefont {Chak~Kui}\ \bibnamefont {Wong}},
  \bibinfo {author} {\bibfnamefont {Pablo}\ \bibnamefont {Llombart}}, \ and\
  \bibinfo {author} {\bibfnamefont {Jonathan~PK}\ \bibnamefont {Doye}},\
  }\bibfield  {title} {\enquote {\bibinfo {title} {How to design an icosahedral
  quasicrystal through directional bonding},}\ }\href@noop {} {\bibfield
  {journal} {\bibinfo  {journal} {Nature}\ }\textbf {\bibinfo {volume} {596}},\
  \bibinfo {pages} {367--371} (\bibinfo {year} {2021})}\BibitemShut {NoStop}%
\bibitem [{\citenamefont {Jeon}\ and\ \citenamefont
  {Lee}(2021{\natexlab{a}})}]{PhysRevResearch.3.013168}%
  \BibitemOpen
  \bibfield  {author} {\bibinfo {author} {\bibfnamefont {Junmo}\ \bibnamefont
  {Jeon}}\ and\ \bibinfo {author} {\bibfnamefont {SungBin}\ \bibnamefont
  {Lee}},\ }\bibfield  {title} {\enquote {\bibinfo {title} {Topological
  critical states and anomalous electronic transmittance in one-dimensional
  quasicrystals},}\ }\href@noop {} {\bibfield  {journal} {\bibinfo  {journal}
  {Phys. Rev. Research}\ }\textbf {\bibinfo {volume} {3}},\ \bibinfo {pages}
  {013168} (\bibinfo {year} {2021}{\natexlab{a}})}\BibitemShut {NoStop}%
\bibitem [{\citenamefont {Jeon}\ and\ \citenamefont
  {Lee}(2020)}]{jeon2020phonon}%
  \BibitemOpen
  \bibfield  {author} {\bibinfo {author} {\bibfnamefont {Junmo}\ \bibnamefont
  {Jeon}}\ and\ \bibinfo {author} {\bibfnamefont {SungBin}\ \bibnamefont
  {Lee}},\ }\bibfield  {title} {\enquote {\bibinfo {title} {Phonon
  transmittance of one dimensional quasicrystals},}\ }\href@noop {} {\bibfield
  {journal} {\bibinfo  {journal} {arXiv preprint arXiv:2009.12378}\ } (\bibinfo
  {year} {2020})}\BibitemShut {NoStop}%
\bibitem [{\citenamefont {Jeon}\ \emph {et~al.}(2021)\citenamefont {Jeon},
  \citenamefont {Park},\ and\ \citenamefont {Lee}}]{jeon2021length}%
  \BibitemOpen
  \bibfield  {author} {\bibinfo {author} {\bibfnamefont {Junmo}\ \bibnamefont
  {Jeon}}, \bibinfo {author} {\bibfnamefont {Moon~Jip}\ \bibnamefont {Park}}, \
  and\ \bibinfo {author} {\bibfnamefont {SungBin}\ \bibnamefont {Lee}},\
  }\bibfield  {title} {\enquote {\bibinfo {title} {Length scale formation in
  the landau levels of quasicrystals},}\ }\href@noop {} {\bibfield  {journal}
  {\bibinfo  {journal} {arXiv preprint arXiv:2106.07782}\ } (\bibinfo {year}
  {2021})}\BibitemShut {NoStop}%
\bibitem [{\citenamefont {Jeon}\ and\ \citenamefont
  {Lee}(2021{\natexlab{b}})}]{jeon2021pattern}%
  \BibitemOpen
  \bibfield  {author} {\bibinfo {author} {\bibfnamefont {Junmo}\ \bibnamefont
  {Jeon}}\ and\ \bibinfo {author} {\bibfnamefont {SungBin}\ \bibnamefont
  {Lee}},\ }\bibfield  {title} {\enquote {\bibinfo {title} {Pattern-dependent
  proximity effect and majorana edge mode in one-dimensional quasicrystals},}\
  }\href@noop {} {\bibfield  {journal} {\bibinfo  {journal} {arXiv preprint
  arXiv:2108.02212}\ } (\bibinfo {year} {2021}{\natexlab{b}})}\BibitemShut
  {NoStop}%
\bibitem [{\citenamefont {Stadnik}(2012)}]{stadnik2012physical}%
  \BibitemOpen
  \bibfield  {author} {\bibinfo {author} {\bibfnamefont {Zbigniew~M}\
  \bibnamefont {Stadnik}},\ }\href@noop {} {\emph {\bibinfo {title} {Physical
  properties of quasicrystals}}},\ Vol.\ \bibinfo {volume} {126}\ (\bibinfo
  {publisher} {Springer Science \& Business Media},\ \bibinfo {year}
  {2012})\BibitemShut {NoStop}%
\bibitem [{\citenamefont {Schwabe}\ \emph {et~al.}(1999)\citenamefont
  {Schwabe}, \citenamefont {Kasner},\ and\ \citenamefont
  {B{\"o}ttger}}]{schwabe1999influence}%
  \BibitemOpen
  \bibfield  {author} {\bibinfo {author} {\bibfnamefont {H}~\bibnamefont
  {Schwabe}}, \bibinfo {author} {\bibfnamefont {G}~\bibnamefont {Kasner}}, \
  and\ \bibinfo {author} {\bibfnamefont {H}~\bibnamefont {B{\"o}ttger}},\
  }\bibfield  {title} {\enquote {\bibinfo {title} {Influence of phason flips on
  electronical properties of quasicrystalline model systems},}\ }\href@noop {}
  {\bibfield  {journal} {\bibinfo  {journal} {Physical Review B}\ }\textbf
  {\bibinfo {volume} {59}},\ \bibinfo {pages} {861} (\bibinfo {year}
  {1999})}\BibitemShut {NoStop}%
\bibitem [{\citenamefont {Kohmoto}\ \emph {et~al.}(1987)\citenamefont
  {Kohmoto}, \citenamefont {Sutherland},\ and\ \citenamefont
  {Tang}}]{kohmoto1987critical}%
  \BibitemOpen
  \bibfield  {author} {\bibinfo {author} {\bibfnamefont {Mahito}\ \bibnamefont
  {Kohmoto}}, \bibinfo {author} {\bibfnamefont {Bill}\ \bibnamefont
  {Sutherland}}, \ and\ \bibinfo {author} {\bibfnamefont {Chao}\ \bibnamefont
  {Tang}},\ }\bibfield  {title} {\enquote {\bibinfo {title} {Critical wave
  functions and a cantor-set spectrum of a one-dimensional quasicrystal
  model},}\ }\href@noop {} {\bibfield  {journal} {\bibinfo  {journal} {Physical
  Review B}\ }\textbf {\bibinfo {volume} {35}},\ \bibinfo {pages} {1020}
  (\bibinfo {year} {1987})}\BibitemShut {NoStop}%
\bibitem [{\citenamefont {Mac\'e}\ \emph {et~al.}(2017)\citenamefont {Mac\'e},
  \citenamefont {Jagannathan}, \citenamefont {Kalugin}, \citenamefont
  {Mosseri},\ and\ \citenamefont {Pi\'echon}}]{mace2017critical}%
  \BibitemOpen
  \bibfield  {author} {\bibinfo {author} {\bibfnamefont {Nicolas}\ \bibnamefont
  {Mac\'e}}, \bibinfo {author} {\bibfnamefont {Anuradha}\ \bibnamefont
  {Jagannathan}}, \bibinfo {author} {\bibfnamefont {Pavel}\ \bibnamefont
  {Kalugin}}, \bibinfo {author} {\bibfnamefont {R\'emy}\ \bibnamefont
  {Mosseri}}, \ and\ \bibinfo {author} {\bibfnamefont {Fr\'ed\'eric}\
  \bibnamefont {Pi\'echon}},\ }\bibfield  {title} {\enquote {\bibinfo {title}
  {Critical eigenstates and their properties in one- and two-dimensional
  quasicrystals},}\ }\href@noop {} {\bibfield  {journal} {\bibinfo  {journal}
  {Phys. Rev. B}\ }\textbf {\bibinfo {volume} {96}},\ \bibinfo {pages} {045138}
  (\bibinfo {year} {2017})}\BibitemShut {NoStop}%
\bibitem [{\citenamefont {Vekilov}\ \emph {et~al.}(2000)\citenamefont
  {Vekilov}, \citenamefont {Isaev},\ and\ \citenamefont
  {Arslanov}}]{vekilov2000influence}%
  \BibitemOpen
  \bibfield  {author} {\bibinfo {author} {\bibfnamefont {Yu~Kh}\ \bibnamefont
  {Vekilov}}, \bibinfo {author} {\bibfnamefont {EI}~\bibnamefont {Isaev}}, \
  and\ \bibinfo {author} {\bibfnamefont {SF}~\bibnamefont {Arslanov}},\
  }\bibfield  {title} {\enquote {\bibinfo {title} {Influence of phason flips,
  magnetic field, and chemical disorder on the localization of electronic
  states in an icosahedral quasicrystal},}\ }\href@noop {} {\bibfield
  {journal} {\bibinfo  {journal} {Physical Review B}\ }\textbf {\bibinfo
  {volume} {62}},\ \bibinfo {pages} {14040} (\bibinfo {year}
  {2000})}\BibitemShut {NoStop}%
\bibitem [{\citenamefont {Bandres}\ \emph {et~al.}(2016)\citenamefont
  {Bandres}, \citenamefont {Rechtsman},\ and\ \citenamefont
  {Segev}}]{bandres2016topological}%
  \BibitemOpen
  \bibfield  {author} {\bibinfo {author} {\bibfnamefont {Miguel~A}\
  \bibnamefont {Bandres}}, \bibinfo {author} {\bibfnamefont {Mikael~C}\
  \bibnamefont {Rechtsman}}, \ and\ \bibinfo {author} {\bibfnamefont
  {Mordechai}\ \bibnamefont {Segev}},\ }\bibfield  {title} {\enquote {\bibinfo
  {title} {Topological photonic quasicrystals: Fractal topological spectrum and
  protected transport},}\ }\href@noop {} {\bibfield  {journal} {\bibinfo
  {journal} {Physical Review X}\ }\textbf {\bibinfo {volume} {6}},\ \bibinfo
  {pages} {011016} (\bibinfo {year} {2016})}\BibitemShut {NoStop}%
\bibitem [{\citenamefont {Freedman}\ \emph {et~al.}(2006)\citenamefont
  {Freedman}, \citenamefont {Bartal}, \citenamefont {Segev}, \citenamefont
  {Lifshitz}, \citenamefont {Christodoulides},\ and\ \citenamefont
  {Fleischer}}]{freedman2006wave}%
  \BibitemOpen
  \bibfield  {author} {\bibinfo {author} {\bibfnamefont {Barak}\ \bibnamefont
  {Freedman}}, \bibinfo {author} {\bibfnamefont {Guy}\ \bibnamefont {Bartal}},
  \bibinfo {author} {\bibfnamefont {Mordechai}\ \bibnamefont {Segev}}, \bibinfo
  {author} {\bibfnamefont {Ron}\ \bibnamefont {Lifshitz}}, \bibinfo {author}
  {\bibfnamefont {Demetrios~N}\ \bibnamefont {Christodoulides}}, \ and\
  \bibinfo {author} {\bibfnamefont {Jason~W}\ \bibnamefont {Fleischer}},\
  }\bibfield  {title} {\enquote {\bibinfo {title} {Wave and defect dynamics in
  nonlinear photonic quasicrystals},}\ }\href@noop {} {\bibfield  {journal}
  {\bibinfo  {journal} {Nature}\ }\textbf {\bibinfo {volume} {440}},\ \bibinfo
  {pages} {1166--1169} (\bibinfo {year} {2006})}\BibitemShut {NoStop}%
\bibitem [{\citenamefont {Yu}\ \emph {et~al.}(2019)\citenamefont {Yu},
  \citenamefont {Wu}, \citenamefont {Zhan}, \citenamefont {Katsnelson},\ and\
  \citenamefont {Yuan}}]{yu2019dodecagonal}%
  \BibitemOpen
  \bibfield  {author} {\bibinfo {author} {\bibfnamefont {Guodong}\ \bibnamefont
  {Yu}}, \bibinfo {author} {\bibfnamefont {Zewen}\ \bibnamefont {Wu}}, \bibinfo
  {author} {\bibfnamefont {Zhen}\ \bibnamefont {Zhan}}, \bibinfo {author}
  {\bibfnamefont {Mikhail~I}\ \bibnamefont {Katsnelson}}, \ and\ \bibinfo
  {author} {\bibfnamefont {Shengjun}\ \bibnamefont {Yuan}},\ }\bibfield
  {title} {\enquote {\bibinfo {title} {Dodecagonal bilayer graphene
  quasicrystal and its approximants},}\ }\href@noop {} {\bibfield  {journal}
  {\bibinfo  {journal} {npj Computational Materials}\ }\textbf {\bibinfo
  {volume} {5}},\ \bibinfo {pages} {1--10} (\bibinfo {year}
  {2019})}\BibitemShut {NoStop}%
\bibitem [{\citenamefont {Fang}\ \emph {et~al.}(2018)\citenamefont {Fang},
  \citenamefont {Paduroiu}, \citenamefont {Hammock},\ and\ \citenamefont
  {Irwin}}]{cryst8110416}%
  \BibitemOpen
  \bibfield  {author} {\bibinfo {author} {\bibfnamefont {Fang}\ \bibnamefont
  {Fang}}, \bibinfo {author} {\bibfnamefont {Sinziana}\ \bibnamefont
  {Paduroiu}}, \bibinfo {author} {\bibfnamefont {Dugan}\ \bibnamefont
  {Hammock}}, \ and\ \bibinfo {author} {\bibfnamefont {Klee}\ \bibnamefont
  {Irwin}},\ }\bibfield  {title} {\enquote {\bibinfo {title} {Non-local game of
  life in 2d quasicrystals},}\ }\href@noop {} {\bibfield  {journal} {\bibinfo
  {journal} {Crystals}\ }\textbf {\bibinfo {volume} {8}} (\bibinfo {year}
  {2018})}\BibitemShut {NoStop}%
\bibitem [{\citenamefont {Arai}\ \emph {et~al.}(1988)\citenamefont {Arai},
  \citenamefont {Tokihiro}, \citenamefont {Fujiwara},\ and\ \citenamefont
  {Kohmoto}}]{arai1988strictly}%
  \BibitemOpen
  \bibfield  {author} {\bibinfo {author} {\bibfnamefont {Masao}\ \bibnamefont
  {Arai}}, \bibinfo {author} {\bibfnamefont {Tetsuji}\ \bibnamefont
  {Tokihiro}}, \bibinfo {author} {\bibfnamefont {Takeo}\ \bibnamefont
  {Fujiwara}}, \ and\ \bibinfo {author} {\bibfnamefont {Mahito}\ \bibnamefont
  {Kohmoto}},\ }\bibfield  {title} {\enquote {\bibinfo {title} {Strictly
  localized states on a two-dimensional penrose lattice},}\ }\href@noop {}
  {\bibfield  {journal} {\bibinfo  {journal} {Physical Review B}\ }\textbf
  {\bibinfo {volume} {38}},\ \bibinfo {pages} {1621} (\bibinfo {year}
  {1988})}\BibitemShut {NoStop}%
\bibitem [{\citenamefont {Kim}\ \emph {et~al.}(2007)\citenamefont {Kim},
  \citenamefont {Kee},\ and\ \citenamefont {Lee}}]{kim2007novel}%
  \BibitemOpen
  \bibfield  {author} {\bibinfo {author} {\bibfnamefont {Soan}\ \bibnamefont
  {Kim}}, \bibinfo {author} {\bibfnamefont {Chul-Sik}\ \bibnamefont {Kee}}, \
  and\ \bibinfo {author} {\bibfnamefont {Jongmin}\ \bibnamefont {Lee}},\
  }\bibfield  {title} {\enquote {\bibinfo {title} {Novel optical properties of
  six-fold symmetric photonic quasicrystal fibers},}\ }\href@noop {} {\bibfield
   {journal} {\bibinfo  {journal} {Optics Express}\ }\textbf {\bibinfo {volume}
  {15}},\ \bibinfo {pages} {13221--13226} (\bibinfo {year} {2007})}\BibitemShut
  {NoStop}%
\bibitem [{\citenamefont {{O}ktel}(2021)}]{oktel2021strictly}%
  \BibitemOpen
  \bibfield  {author} {\bibinfo {author} {\bibfnamefont {M.{O}.}\ \bibnamefont
  {{O}ktel}},\ }\bibfield  {title} {\enquote {\bibinfo {title} {Strictly
  localized states in the octagonal ammann-beenker quasicrystal},}\ }\href@noop
  {} {\bibfield  {journal} {\bibinfo  {journal} {Physical Review B}\ }\textbf
  {\bibinfo {volume} {104}},\ \bibinfo {pages} {014204} (\bibinfo {year}
  {2021})}\BibitemShut {NoStop}%
\bibitem [{\citenamefont {Hammock}\ \emph {et~al.}(2018)\citenamefont
  {Hammock}, \citenamefont {Fang},\ and\ \citenamefont {Irwin}}]{cryst8100370}%
  \BibitemOpen
  \bibfield  {author} {\bibinfo {author} {\bibfnamefont {Dugan}\ \bibnamefont
  {Hammock}}, \bibinfo {author} {\bibfnamefont {Fang}\ \bibnamefont {Fang}}, \
  and\ \bibinfo {author} {\bibfnamefont {Klee}\ \bibnamefont {Irwin}},\
  }\bibfield  {title} {\enquote {\bibinfo {title} {Quasicrystal tilings in
  three dimensions and their empires},}\ }\href@noop {} {\bibfield  {journal}
  {\bibinfo  {journal} {Crystals}\ }\textbf {\bibinfo {volume} {8}} (\bibinfo
  {year} {2018})}\BibitemShut {NoStop}%
\bibitem [{\citenamefont {Fang}\ \emph {et~al.}(2017)\citenamefont {Fang},
  \citenamefont {Hammock},\ and\ \citenamefont {Irwin}}]{cryst7100304}%
  \BibitemOpen
  \bibfield  {author} {\bibinfo {author} {\bibfnamefont {Fang}\ \bibnamefont
  {Fang}}, \bibinfo {author} {\bibfnamefont {Dugan}\ \bibnamefont {Hammock}}, \
  and\ \bibinfo {author} {\bibfnamefont {Klee}\ \bibnamefont {Irwin}},\
  }\bibfield  {title} {\enquote {\bibinfo {title} {Methods for calculating
  empires in quasicrystals},}\ }\href@noop {} {\bibfield  {journal} {\bibinfo
  {journal} {Crystals}\ }\textbf {\bibinfo {volume} {7}} (\bibinfo {year}
  {2017})}\BibitemShut {NoStop}%
\bibitem [{\citenamefont {de~Boissieu}(2012)}]{de2012phonons}%
  \BibitemOpen
  \bibfield  {author} {\bibinfo {author} {\bibfnamefont {Marc}\ \bibnamefont
  {de~Boissieu}},\ }\bibfield  {title} {\enquote {\bibinfo {title} {Phonons,
  phasons and atomic dynamics in quasicrystals},}\ }\href@noop {} {\bibfield
  {journal} {\bibinfo  {journal} {Chemical Society Reviews}\ }\textbf {\bibinfo
  {volume} {41}},\ \bibinfo {pages} {6778--6786} (\bibinfo {year}
  {2012})}\BibitemShut {NoStop}%
\bibitem [{\citenamefont {Homes}\ \emph {et~al.}(1991)\citenamefont {Homes},
  \citenamefont {Timusk}, \citenamefont {Wu}, \citenamefont {Altounian},
  \citenamefont {Sahnoune},\ and\ \citenamefont
  {Str{\"o}m-Olsen}}]{homes1991optical}%
  \BibitemOpen
  \bibfield  {author} {\bibinfo {author} {\bibfnamefont {CC}~\bibnamefont
  {Homes}}, \bibinfo {author} {\bibfnamefont {T}~\bibnamefont {Timusk}},
  \bibinfo {author} {\bibfnamefont {X}~\bibnamefont {Wu}}, \bibinfo {author}
  {\bibfnamefont {Z}~\bibnamefont {Altounian}}, \bibinfo {author}
  {\bibfnamefont {A}~\bibnamefont {Sahnoune}}, \ and\ \bibinfo {author}
  {\bibfnamefont {JO}~\bibnamefont {Str{\"o}m-Olsen}},\ }\bibfield  {title}
  {\enquote {\bibinfo {title} {Optical conductivity of the stable icosahedral
  quasicrystal al 63.5 cu 24.5 fe 12},}\ }\href@noop {} {\bibfield  {journal}
  {\bibinfo  {journal} {Physical review letters}\ }\textbf {\bibinfo {volume}
  {67}},\ \bibinfo {pages} {2694} (\bibinfo {year} {1991})}\BibitemShut
  {NoStop}%
\bibitem [{\citenamefont {de~Boissieu}(2011)}]{de2011phonons}%
  \BibitemOpen
  \bibfield  {author} {\bibinfo {author} {\bibfnamefont {Marc}\ \bibnamefont
  {de~Boissieu}},\ }\bibfield  {title} {\enquote {\bibinfo {title} {Phonons and
  phasons in icosahedral quasicrystals},}\ }\href@noop {} {\bibfield  {journal}
  {\bibinfo  {journal} {Israel Journal of Chemistry}\ }\textbf {\bibinfo
  {volume} {51}},\ \bibinfo {pages} {1292--1303} (\bibinfo {year}
  {2011})}\BibitemShut {NoStop}%
\bibitem [{\citenamefont {Bistritzer}\ and\ \citenamefont
  {MacDonald}(2011)}]{bistritzer2011moire}%
  \BibitemOpen
  \bibfield  {author} {\bibinfo {author} {\bibfnamefont {R}~\bibnamefont
  {Bistritzer}}\ and\ \bibinfo {author} {\bibfnamefont {AH}~\bibnamefont
  {MacDonald}},\ }\bibfield  {title} {\enquote {\bibinfo {title} {Moir{\'e}
  butterflies in twisted bilayer graphene},}\ }\href@noop {} {\bibfield
  {journal} {\bibinfo  {journal} {Physical Review B}\ }\textbf {\bibinfo
  {volume} {84}},\ \bibinfo {pages} {035440} (\bibinfo {year}
  {2011})}\BibitemShut {NoStop}%
\bibitem [{\citenamefont {Nemec}\ and\ \citenamefont
  {Cuniberti}(2007)}]{nemec2007hofstadter}%
  \BibitemOpen
  \bibfield  {author} {\bibinfo {author} {\bibfnamefont {Norbert}\ \bibnamefont
  {Nemec}}\ and\ \bibinfo {author} {\bibfnamefont {Gianaurelio}\ \bibnamefont
  {Cuniberti}},\ }\bibfield  {title} {\enquote {\bibinfo {title} {Hofstadter
  butterflies of bilayer graphene},}\ }\href@noop {} {\bibfield  {journal}
  {\bibinfo  {journal} {Physical Review B}\ }\textbf {\bibinfo {volume} {75}},\
  \bibinfo {pages} {201404} (\bibinfo {year} {2007})}\BibitemShut {NoStop}%
\bibitem [{\citenamefont {Pezzini}\ \emph {et~al.}(2020)\citenamefont
  {Pezzini}, \citenamefont {Miseikis}, \citenamefont {Piccinini}, \citenamefont
  {Forti}, \citenamefont {Pace}, \citenamefont {Engelke}, \citenamefont
  {Rossella}, \citenamefont {Watanabe}, \citenamefont {Taniguchi},
  \citenamefont {Kim} \emph {et~al.}}]{pezzini202030}%
  \BibitemOpen
  \bibfield  {author} {\bibinfo {author} {\bibfnamefont {Sergio}\ \bibnamefont
  {Pezzini}}, \bibinfo {author} {\bibfnamefont {Vaidotas}\ \bibnamefont
  {Miseikis}}, \bibinfo {author} {\bibfnamefont {Giulia}\ \bibnamefont
  {Piccinini}}, \bibinfo {author} {\bibfnamefont {Stiven}\ \bibnamefont
  {Forti}}, \bibinfo {author} {\bibfnamefont {Simona}\ \bibnamefont {Pace}},
  \bibinfo {author} {\bibfnamefont {Rebecca}\ \bibnamefont {Engelke}}, \bibinfo
  {author} {\bibfnamefont {Francesco}\ \bibnamefont {Rossella}}, \bibinfo
  {author} {\bibfnamefont {Kenji}\ \bibnamefont {Watanabe}}, \bibinfo {author}
  {\bibfnamefont {Takashi}\ \bibnamefont {Taniguchi}}, \bibinfo {author}
  {\bibfnamefont {Philip}\ \bibnamefont {Kim}},  \emph {et~al.},\ }\bibfield
  {title} {\enquote {\bibinfo {title} {30-twisted bilayer graphene
  quasicrystals from chemical vapor deposition},}\ }\href@noop {} {\bibfield
  {journal} {\bibinfo  {journal} {Nano letters}\ }\textbf {\bibinfo {volume}
  {20}},\ \bibinfo {pages} {3313--3319} (\bibinfo {year} {2020})}\BibitemShut
  {NoStop}%
\bibitem [{\citenamefont {Reinhardt}\ \emph {et~al.}(2013)\citenamefont
  {Reinhardt}, \citenamefont {Romano},\ and\ \citenamefont
  {Doye}}]{reinhardt2013computing}%
  \BibitemOpen
  \bibfield  {author} {\bibinfo {author} {\bibfnamefont {Aleks}\ \bibnamefont
  {Reinhardt}}, \bibinfo {author} {\bibfnamefont {Flavio}\ \bibnamefont
  {Romano}}, \ and\ \bibinfo {author} {\bibfnamefont {Jonathan~PK}\
  \bibnamefont {Doye}},\ }\bibfield  {title} {\enquote {\bibinfo {title}
  {Computing phase diagrams for a quasicrystal-forming patchy-particle
  system},}\ }\href@noop {} {\bibfield  {journal} {\bibinfo  {journal}
  {Physical review letters}\ }\textbf {\bibinfo {volume} {110}},\ \bibinfo
  {pages} {255503} (\bibinfo {year} {2013})}\BibitemShut {NoStop}%
\bibitem [{\citenamefont {Tracey}\ \emph {et~al.}(2021)\citenamefont {Tracey},
  \citenamefont {Noya},\ and\ \citenamefont {Doye}}]{tracey2021programming}%
  \BibitemOpen
  \bibfield  {author} {\bibinfo {author} {\bibfnamefont {Daniel~F}\
  \bibnamefont {Tracey}}, \bibinfo {author} {\bibfnamefont {Eva~G}\
  \bibnamefont {Noya}}, \ and\ \bibinfo {author} {\bibfnamefont {Jonathan~PK}\
  \bibnamefont {Doye}},\ }\bibfield  {title} {\enquote {\bibinfo {title}
  {Programming patchy particles to form three-dimensional dodecagonal
  quasicrystals},}\ }\href@noop {} {\bibfield  {journal} {\bibinfo  {journal}
  {The Journal of Chemical Physics}\ }\textbf {\bibinfo {volume} {154}},\
  \bibinfo {pages} {194505} (\bibinfo {year} {2021})}\BibitemShut {NoStop}%
\bibitem [{\citenamefont {Senechal}(1996)}]{senechal1996quasicrystals}%
  \BibitemOpen
  \bibfield  {author} {\bibinfo {author} {\bibfnamefont {M.}~\bibnamefont
  {Senechal}},\ }\href@noop {} {\emph {\bibinfo {title} {Quasicrystals and
  Geometry}}}\ (\bibinfo  {publisher} {Cambridge University Press},\ \bibinfo
  {year} {1996})\BibitemShut {NoStop}%
\bibitem [{\citenamefont {Arag{\'o}n}\ \emph {et~al.}(2019)\citenamefont
  {Arag{\'o}n}, \citenamefont {Naumis},\ and\ \citenamefont
  {G{\'o}mez-Rodr{\'\i}guez}}]{aragon2019twisted}%
  \BibitemOpen
  \bibfield  {author} {\bibinfo {author} {\bibfnamefont {Jos{\'e}~L}\
  \bibnamefont {Arag{\'o}n}}, \bibinfo {author} {\bibfnamefont {Gerardo~G}\
  \bibnamefont {Naumis}}, \ and\ \bibinfo {author} {\bibfnamefont {Alfredo}\
  \bibnamefont {G{\'o}mez-Rodr{\'\i}guez}},\ }\bibfield  {title} {\enquote
  {\bibinfo {title} {Twisted graphene bilayers and quasicrystals: A cut and
  projection approach},}\ }\href@noop {} {\bibfield  {journal} {\bibinfo
  {journal} {Crystals}\ }\textbf {\bibinfo {volume} {9}},\ \bibinfo {pages}
  {519} (\bibinfo {year} {2019})}\BibitemShut {NoStop}%
\bibitem [{\citenamefont {Indelicato}\ \emph {et~al.}(2012)\citenamefont
  {Indelicato}, \citenamefont {Keef}, \citenamefont {Cermelli}, \citenamefont
  {Salthouse}, \citenamefont {Twarock},\ and\ \citenamefont
  {Zanzotto}}]{indelicato2012structural}%
  \BibitemOpen
  \bibfield  {author} {\bibinfo {author} {\bibfnamefont {Giuliana}\
  \bibnamefont {Indelicato}}, \bibinfo {author} {\bibfnamefont {Tom}\
  \bibnamefont {Keef}}, \bibinfo {author} {\bibfnamefont {Paolo}\ \bibnamefont
  {Cermelli}}, \bibinfo {author} {\bibfnamefont {David~G}\ \bibnamefont
  {Salthouse}}, \bibinfo {author} {\bibfnamefont {Reidun}\ \bibnamefont
  {Twarock}}, \ and\ \bibinfo {author} {\bibfnamefont {Giovanni}\ \bibnamefont
  {Zanzotto}},\ }\bibfield  {title} {\enquote {\bibinfo {title} {Structural
  transformations in quasicrystals induced by higher dimensional lattice
  transitions},}\ }\href@noop {} {\bibfield  {journal} {\bibinfo  {journal}
  {Proceedings of the Royal Society A: Mathematical, Physical and Engineering
  Sciences}\ }\textbf {\bibinfo {volume} {468}},\ \bibinfo {pages} {1452--1471}
  (\bibinfo {year} {2012})}\BibitemShut {NoStop}%
\bibitem [{\citenamefont {Jagannathan}\ and\ \citenamefont
  {Duneau}(2014)}]{jagannathan2014eightfold}%
  \BibitemOpen
  \bibfield  {author} {\bibinfo {author} {\bibfnamefont {Anuradha}\
  \bibnamefont {Jagannathan}}\ and\ \bibinfo {author} {\bibfnamefont {Michel}\
  \bibnamefont {Duneau}},\ }\bibfield  {title} {\enquote {\bibinfo {title} {An
  eightfold optical quasicrystal with cold atoms},}\ }\href@noop {} {\bibfield
  {journal} {\bibinfo  {journal} {EPL (Europhysics Letters)}\ }\textbf
  {\bibinfo {volume} {104}},\ \bibinfo {pages} {66003} (\bibinfo {year}
  {2014})}\BibitemShut {NoStop}%
\bibitem [{\citenamefont {Corcovilos}\ and\ \citenamefont
  {Mittal}(2019)}]{corcovilos2019two}%
  \BibitemOpen
  \bibfield  {author} {\bibinfo {author} {\bibfnamefont {Theodore~A}\
  \bibnamefont {Corcovilos}}\ and\ \bibinfo {author} {\bibfnamefont {Jahnavee}\
  \bibnamefont {Mittal}},\ }\bibfield  {title} {\enquote {\bibinfo {title}
  {Two-dimensional optical quasicrystal potentials for ultracold atom
  experiments},}\ }\href@noop {} {\bibfield  {journal} {\bibinfo  {journal}
  {Applied optics}\ }\textbf {\bibinfo {volume} {58}},\ \bibinfo {pages}
  {2256--2263} (\bibinfo {year} {2019})}\BibitemShut {NoStop}%
\bibitem [{\citenamefont {Sanchez-Palencia}\ and\ \citenamefont
  {Santos}(2005)}]{sanchez2005bose}%
  \BibitemOpen
  \bibfield  {author} {\bibinfo {author} {\bibfnamefont {Laurent}\ \bibnamefont
  {Sanchez-Palencia}}\ and\ \bibinfo {author} {\bibfnamefont {Luis}\
  \bibnamefont {Santos}},\ }\bibfield  {title} {\enquote {\bibinfo {title}
  {Bose-einstein condensates in optical quasicrystal lattices},}\ }\href@noop
  {} {\bibfield  {journal} {\bibinfo  {journal} {Physical Review A}\ }\textbf
  {\bibinfo {volume} {72}},\ \bibinfo {pages} {053607} (\bibinfo {year}
  {2005})}\BibitemShut {NoStop}%
\bibitem [{\citenamefont {Viebahn}\ \emph {et~al.}(2019)\citenamefont
  {Viebahn}, \citenamefont {Sbroscia}, \citenamefont {Carter}, \citenamefont
  {Yu},\ and\ \citenamefont {Schneider}}]{viebahn2019matter}%
  \BibitemOpen
  \bibfield  {author} {\bibinfo {author} {\bibfnamefont {Konrad}\ \bibnamefont
  {Viebahn}}, \bibinfo {author} {\bibfnamefont {Matteo}\ \bibnamefont
  {Sbroscia}}, \bibinfo {author} {\bibfnamefont {Edward}\ \bibnamefont
  {Carter}}, \bibinfo {author} {\bibfnamefont {Jr-Chiun}\ \bibnamefont {Yu}}, \
  and\ \bibinfo {author} {\bibfnamefont {Ulrich}\ \bibnamefont {Schneider}},\
  }\bibfield  {title} {\enquote {\bibinfo {title} {Matter-wave diffraction from
  a quasicrystalline optical lattice},}\ }\href@noop {} {\bibfield  {journal}
  {\bibinfo  {journal} {Physical review letters}\ }\textbf {\bibinfo {volume}
  {122}},\ \bibinfo {pages} {110404} (\bibinfo {year} {2019})}\BibitemShut
  {NoStop}%
\bibitem [{\citenamefont {Sbroscia}\ \emph {et~al.}(2020)\citenamefont
  {Sbroscia}, \citenamefont {Viebahn}, \citenamefont {Carter}, \citenamefont
  {Yu}, \citenamefont {Gaunt},\ and\ \citenamefont
  {Schneider}}]{sbroscia2020observing}%
  \BibitemOpen
  \bibfield  {author} {\bibinfo {author} {\bibfnamefont {Matteo}\ \bibnamefont
  {Sbroscia}}, \bibinfo {author} {\bibfnamefont {Konrad}\ \bibnamefont
  {Viebahn}}, \bibinfo {author} {\bibfnamefont {Edward}\ \bibnamefont
  {Carter}}, \bibinfo {author} {\bibfnamefont {Jr-Chiun}\ \bibnamefont {Yu}},
  \bibinfo {author} {\bibfnamefont {Alexander}\ \bibnamefont {Gaunt}}, \ and\
  \bibinfo {author} {\bibfnamefont {Ulrich}\ \bibnamefont {Schneider}},\
  }\bibfield  {title} {\enquote {\bibinfo {title} {Observing localization in a
  2d quasicrystalline optical lattice},}\ }\href@noop {} {\bibfield  {journal}
  {\bibinfo  {journal} {Physical Review Letters}\ }\textbf {\bibinfo {volume}
  {125}},\ \bibinfo {pages} {200604} (\bibinfo {year} {2020})}\BibitemShut
  {NoStop}%
\bibitem [{\citenamefont {Zhao}\ \emph {et~al.}(2013)\citenamefont {Zhao},
  \citenamefont {Yu},\ and\ \citenamefont {Xu}}]{zhao2013hamiltonian}%
  \BibitemOpen
  \bibfield  {author} {\bibinfo {author} {\bibfnamefont {Liu}\ \bibnamefont
  {Zhao}}, \bibinfo {author} {\bibfnamefont {Pengfei}\ \bibnamefont {Yu}}, \
  and\ \bibinfo {author} {\bibfnamefont {Wei}\ \bibnamefont {Xu}},\ }\bibfield
  {title} {\enquote {\bibinfo {title} {Hamiltonian description of singular
  lagrangian systems with spontaneously broken time translation symmetry},}\
  }\href@noop {} {\bibfield  {journal} {\bibinfo  {journal} {Modern Physics
  Letters A}\ }\textbf {\bibinfo {volume} {28}},\ \bibinfo {pages} {1350002}
  (\bibinfo {year} {2013})}\BibitemShut {NoStop}%
\bibitem [{\citenamefont {Russomanno}\ \emph {et~al.}(2017)\citenamefont
  {Russomanno}, \citenamefont {Iemini}, \citenamefont {Dalmonte},\ and\
  \citenamefont {Fazio}}]{russomanno2017floquet}%
  \BibitemOpen
  \bibfield  {author} {\bibinfo {author} {\bibfnamefont {Angelo}\ \bibnamefont
  {Russomanno}}, \bibinfo {author} {\bibfnamefont {Fernando}\ \bibnamefont
  {Iemini}}, \bibinfo {author} {\bibfnamefont {Marcello}\ \bibnamefont
  {Dalmonte}}, \ and\ \bibinfo {author} {\bibfnamefont {Rosario}\ \bibnamefont
  {Fazio}},\ }\bibfield  {title} {\enquote {\bibinfo {title} {Floquet time
  crystal in the lipkin-meshkov-glick model},}\ }\href@noop {} {\bibfield
  {journal} {\bibinfo  {journal} {Physical Review B}\ }\textbf {\bibinfo
  {volume} {95}},\ \bibinfo {pages} {214307} (\bibinfo {year}
  {2017})}\BibitemShut {NoStop}%
\bibitem [{\citenamefont {Martin}\ \emph {et~al.}(2017)\citenamefont {Martin},
  \citenamefont {Refael},\ and\ \citenamefont
  {Halperin}}]{martin2017topological}%
  \BibitemOpen
  \bibfield  {author} {\bibinfo {author} {\bibfnamefont {Ivar}\ \bibnamefont
  {Martin}}, \bibinfo {author} {\bibfnamefont {Gil}\ \bibnamefont {Refael}}, \
  and\ \bibinfo {author} {\bibfnamefont {Bertrand}\ \bibnamefont {Halperin}},\
  }\bibfield  {title} {\enquote {\bibinfo {title} {Topological frequency
  conversion in strongly driven quantum systems},}\ }\href@noop {} {\bibfield
  {journal} {\bibinfo  {journal} {Physical Review X}\ }\textbf {\bibinfo
  {volume} {7}},\ \bibinfo {pages} {041008} (\bibinfo {year}
  {2017})}\BibitemShut {NoStop}%
\bibitem [{\citenamefont {Rodriguez-Vega}\ \emph {et~al.}(2021)\citenamefont
  {Rodriguez-Vega}, \citenamefont {Carlander}, \citenamefont {Bahri},
  \citenamefont {Lin}, \citenamefont {Sinitsyn},\ and\ \citenamefont
  {Fiete}}]{rodriguez2021real}%
  \BibitemOpen
  \bibfield  {author} {\bibinfo {author} {\bibfnamefont {Martin}\ \bibnamefont
  {Rodriguez-Vega}}, \bibinfo {author} {\bibfnamefont {Ella}\ \bibnamefont
  {Carlander}}, \bibinfo {author} {\bibfnamefont {Adrian}\ \bibnamefont
  {Bahri}}, \bibinfo {author} {\bibfnamefont {Ze-Xun}\ \bibnamefont {Lin}},
  \bibinfo {author} {\bibfnamefont {Nikolai~A}\ \bibnamefont {Sinitsyn}}, \
  and\ \bibinfo {author} {\bibfnamefont {Gregory~A}\ \bibnamefont {Fiete}},\
  }\bibfield  {title} {\enquote {\bibinfo {title} {Real-time simulation of
  light-driven spin chains on quantum computers},}\ }\href@noop {} {\bibfield
  {journal} {\bibinfo  {journal} {arXiv preprint arXiv:2108.05975}\ } (\bibinfo
  {year} {2021})}\BibitemShut {NoStop}%
\bibitem [{\citenamefont {Malz}\ and\ \citenamefont
  {Smith}(2021)}]{malz2021topological}%
  \BibitemOpen
  \bibfield  {author} {\bibinfo {author} {\bibfnamefont {Daniel}\ \bibnamefont
  {Malz}}\ and\ \bibinfo {author} {\bibfnamefont {Adam}\ \bibnamefont
  {Smith}},\ }\bibfield  {title} {\enquote {\bibinfo {title} {Topological
  two-dimensional floquet lattice on a single superconducting qubit},}\
  }\href@noop {} {\bibfield  {journal} {\bibinfo  {journal} {Physical Review
  Letters}\ }\textbf {\bibinfo {volume} {126}},\ \bibinfo {pages} {163602}
  (\bibinfo {year} {2021})}\BibitemShut {NoStop}%
\bibitem [{\citenamefont {Poertner}\ and\ \citenamefont
  {Martin}(2020)}]{poertner2020validity}%
  \BibitemOpen
  \bibfield  {author} {\bibinfo {author} {\bibfnamefont {AN}~\bibnamefont
  {Poertner}}\ and\ \bibinfo {author} {\bibfnamefont {JDD}\ \bibnamefont
  {Martin}},\ }\bibfield  {title} {\enquote {\bibinfo {title} {Validity of
  many-mode floquet theory with commensurate frequencies},}\ }\href@noop {}
  {\bibfield  {journal} {\bibinfo  {journal} {Physical Review A}\ }\textbf
  {\bibinfo {volume} {101}},\ \bibinfo {pages} {032116} (\bibinfo {year}
  {2020})}\BibitemShut {NoStop}%
\bibitem [{\citenamefont {Thomson}\ \emph {et~al.}(2021)\citenamefont
  {Thomson}, \citenamefont {Magano},\ and\ \citenamefont
  {Schir{\'o}}}]{thomson2021flow}%
  \BibitemOpen
  \bibfield  {author} {\bibinfo {author} {\bibfnamefont {Steven}\ \bibnamefont
  {Thomson}}, \bibinfo {author} {\bibfnamefont {Duarte}\ \bibnamefont
  {Magano}}, \ and\ \bibinfo {author} {\bibfnamefont {Marco}\ \bibnamefont
  {Schir{\'o}}},\ }\bibfield  {title} {\enquote {\bibinfo {title} {Flow
  equations for disordered floquet systems},}\ }\href@noop {} {\bibfield
  {journal} {\bibinfo  {journal} {SciPost Physics}\ }\textbf {\bibinfo {volume}
  {11}},\ \bibinfo {pages} {028} (\bibinfo {year} {2021})}\BibitemShut
  {NoStop}%
\bibitem [{\citenamefont {van~der Linden}\ \emph {et~al.}(2012)\citenamefont
  {van~der Linden}, \citenamefont {Doye},\ and\ \citenamefont
  {Louis}}]{van2012formation}%
  \BibitemOpen
  \bibfield  {author} {\bibinfo {author} {\bibfnamefont {Marjolein~N}\
  \bibnamefont {van~der Linden}}, \bibinfo {author} {\bibfnamefont
  {Jonathan~PK}\ \bibnamefont {Doye}}, \ and\ \bibinfo {author} {\bibfnamefont
  {Ard~A}\ \bibnamefont {Louis}},\ }\bibfield  {title} {\enquote {\bibinfo
  {title} {Formation of dodecagonal quasicrystals in two-dimensional systems of
  patchy particles},}\ }\href@noop {} {\bibfield  {journal} {\bibinfo
  {journal} {The Journal of chemical physics}\ }\textbf {\bibinfo {volume}
  {136}},\ \bibinfo {pages} {054904} (\bibinfo {year} {2012})}\BibitemShut
  {NoStop}%
\bibitem [{\citenamefont {Sadoc}\ and\ \citenamefont
  {Mosseri}(1993)}]{sadoc1993e8}%
  \BibitemOpen
  \bibfield  {author} {\bibinfo {author} {\bibfnamefont {JF}~\bibnamefont
  {Sadoc}}\ and\ \bibinfo {author} {\bibfnamefont {R}~\bibnamefont {Mosseri}},\
  }\bibfield  {title} {\enquote {\bibinfo {title} {The e8 lattice and
  quasicrystals: geometry, number theory and quasicrystals},}\ }\href@noop {}
  {\bibfield  {journal} {\bibinfo  {journal} {Journal of Physics A:
  Mathematical and General}\ }\textbf {\bibinfo {volume} {26}},\ \bibinfo
  {pages} {1789} (\bibinfo {year} {1993})}\BibitemShut {NoStop}%
\bibitem [{Note1()}]{Note1}%
  \BibitemOpen
  \bibinfo {note} {Junmo Jeon and SungBin Lee, In preparation}\BibitemShut
  {NoStop}%
\end{thebibliography}%

\end{document}